\documentclass[
reprint,
amssymb, amsmath,
bibnotes,
aps, 
superscriptaddress,
 prb
 ]{revtex4-1}

\usepackage{graphicx,hyperref}
\usepackage{dcolumn}
\usepackage{bm}
\usepackage{braket}
\usepackage{float}
\usepackage{color}

\usepackage{hyperref}
\usepackage{natbib}

\begin{document}
\title{First-principles calculation of anomalous Hall and Nernst conductivity\\by local Berry phase}
\author{Hikaru Sawahata}%
\email[Email: ]{sawahata@cphys.s.kanazawa-u.ac.jp}
\affiliation{Nanomaterials Research Institute (NanoMaRi), Kanazawa University, Kanazawa, Ishikawa 920-1192, Japan}
\affiliation{PKSHA technology, Bunkyo-ku, Tokyo 113-0033, Japan}
\author{Naoya Yamaguchi}
\affiliation{Nanomaterials Research Institute (NanoMaRi), Kanazawa University, Kanazawa, Ishikawa 920-1192, Japan}
\author{Susumu Minami}
\affiliation{Department of Physics, University of Tokyo, Bunkyo-ku, Tokyo 113-0033, Japan}
\author{Fumiyuki Ishii}
\email[Email: ]{ishii@cphys.s.kanazawa-u.ac.jp}
\affiliation{Nanomaterials Research Institute (NanoMaRi), Kanazawa University, Kanazawa, Ishikawa 920-1192, Japan}
\date{\today}%
\begin{abstract}
  In this study, we implemented a finite-difference algorithm for computing anomalous Hall and Nernst conductivity. Based on the expression to evaluate the Berry curvature in an insulating system [J. Phys. Soc. Jpn. {\bf 74} 1674(2005)], we extended the methods to a metallic system. We calculated anomalous Hall conductivity and Nernst conductivity in a two-dimensional ferromagnetic material FeCl$_2$ and three-dimensional ferromagnetic transition metals bcc-Fe, hcp-Co, and fcc-Ni. Our results are comparable to previously reported results computed by Kubo-formula or Wannier representation. To evaluate anomalous Nernst coefficients, the detailed Fermi-energy dependence of the anomalous Hall conductivity is required. Nonetheless, previous methods based on Wannier representation or Kubo-formula have numerical instability due to the ${\bf k}$-space Dirac monopole. The present method will open an efficient thermoelectric material design based on the high-throughput first-principles screening.
\end{abstract}
\maketitle
\section{Introduction}
Anomalous Hall effect (AHE) shows Hall conductivity induced by broken time-reversal symmetry based on spontaneous magnetization\cite{Hall,Nagaosa}. AHE has an extrinsic mechanism, which originates from impurities\cite{JSmit,LBerger} and an intrinsic mechanism induced by the Berry curvature\cite{Berry,Kohmoto,DXiao_RevModPhy_2010}. The Berry curvature is a gauge invariant due to the topology of the electron wave function, and AHE occurs in simple ferromagnetic materials and materials with spin chirality\cite{Ohgushi, Nakatsuji}. Therefore, the relation between anomalous Hall conductivity and magnetic structure in materials is non-trivial. In an insulating system, AHE has attracted much research attention as a topological effect because it is quantized as a Chern number, which is a topological invariant\cite{TKNN, Science}.
\par
Anomalous Nernst effect (ANE), which originates from AHE has attracted renewed interest. ANE generates a transverse voltage from a longitudinal temperature gradient due to a transverse electric conductivity through AHE\cite{DXiao, Nagaosa, DXiao_RevModPhy_2010}.
It can be utilized in developing energy-harvesting technology which may provide a simple lateral structure, higher flexibility, and lower production cost  \cite{Sakuraba,Mizuguchi_SciTechAdvMat_2019}.
Experimental and theoretical studies of ANE have been reported in various magnetic materials\cite{Lee_PRL_2004,Miyasato_PRL_2007,Yong_PRL_2008,Sakuraba_APEX_2013,doi:10.1063/1.4922901,YMizuta_H,Mn3Sn,PhysRevB.96.224415,YMizuta_EuO,Minami_APL_2018,Sakai,Liu2018_NatPhys,Guin2019,Guin_AdvMat_2019,doi:10.1021/acs.nanolett.9b03739,Fe3X_Nature,Minami_PRB_2020,Rifky,Sumida_CommunMat_2020,Yang_PhysRevMat_2020,PhysRevApplied.13.054044,Max_PRL_2020,Taishi_NatCom_2021,TAsaba_SicAdv_2021,BinHe_Joule_2021,Nakamura_PRBL_2021,Taishi_SicAdv_2022}. Among them, theoretical work predicted large AHE and ANE in the Skyrmion crystal, which has a spin chirality\cite{YMizuta_H, YMizuta_EuO}. In addition, topological magnets, such as Mn$_3$Sn\cite{Mn3Sn,PhysRevB.96.224415,Taishi_NatCom_2021}, Co$_2$MnGa\cite{Sakai,Guin2019,Sumida_CommunMat_2020}, Fe$_3$X (X=Al, Ga)\cite{Fe3X_Nature,Minami_PRB_2020}, Co$_3$Sn$_2$S$_2$\cite{Liu2018_NatPhys,Guin_AdvMat_2019,Yang_PhysRevMat_2020}, and UCoAl\cite{TAsaba_SicAdv_2021} are particularly interesting due to their large ANE signal and characteristic low-energy electronics structure including Weyl points.\par
Anomalous Nernst conductivity (transverse thermoelectric conductivity), which indicates anomalous Nernst thermoelectric conversion efficiency can be evaluated from the chemical potential dependence of anomalous Hall conductivity. The intrinsic component of anomalous Hall conductivity $\sigma_{xy}$ can be obtained from the Berry curvature ${\bf F}^{n}({\bf k})$ as\cite{TKNN, Kohmoto}, 
\begin{equation}
  \sigma_{xy}(\mu)=-\frac{e^2}{h}\sum_{n}^{N}\int \frac{d{\bf k}^2}{2\pi} F^{n}_{z}({\bf k})f(\varepsilon_{n {\bf k}}-\mu) \label{eq.AHC}.
\end{equation}
Here, $N, e, h, f, \varepsilon_{n {\bf k}}$ and $\mu$ are the electron occupation number, elementary charge, Planck constant, Fermi-Dirac distribution function, band energy with the band index $n$, wave vector ${\bf k}$, and Fermi energy, respectively.
The Berry curvature ${\bf F}^{n}({\bf k})$ is given as,
\begin{eqnarray}
  {\bf F}^{n}({\bf k}) &=&\nabla \times {\bf A}^{n}, \label{eq.defF} \\
  {\bf A}^{n}({\bf k}) &=& -i \Braket{ u^{n}({\bf k})|\nabla_{{\bf k}}|u^{n}({\bf k})}, \label{eq.defA}
\end{eqnarray}
where ${\bf A}^{n}$ and $u^{n}({\bf k})$ are the Berry connection and the periodic part of the Bloch states, respectively.
In an insulating system, Eq.(\ref{eq.AHC}) should be quantized, and it can be described as follows:
$\sigma_{xy}(\varepsilon_{\rm F})=-e^2C/h\ (C=0, \pm1, \pm2, \cdots).$
Here, $\varepsilon_{\rm F}$ is the Fermi energy, and
the integer $C$ is ``Chern number". Anomalous Nernst conductivity $\alpha_{xy}$ can be calculated from the chemical potential dependence of $\sigma_{xy}$ as follows\cite{YMizuta_JPS}:
\begin{equation}
\alpha_{xy}(\mu,T) =\frac{1}{e}\int d\varepsilon\sigma_{xy}(\varepsilon)|_{T=0}\frac{\varepsilon - \mu}{T} \left(-\frac{\partial f(\mu)}{\partial \varepsilon}\right).
\end{equation}
An efficient and simple method to evaluate the chemical potential dependence of $\sigma_{xy}$ is required for the design of thermoelectric materials based on high-throughput first-principles screening. 
Previous studies were mainly performed by evaluating the off-diagonal Hall conductivities with Wannier representation\cite{Vanderbilt, Lopez, Tsirkin} or the Kubo formula\cite{YYao}. 
The former method was implemented in the Wannier90 code\cite{Wannier90} and has been widely used in conjunction with many first-principles electronic structure packages. However, some empirical and technical procedures such as choice of bases and energy window range are required to construct Wannier functions, which are material dependent.
\par

In this paper, we introduce an efficient method for calculating anomalous Hall conductivity $\sigma_{xy}$ and anomalous Nernst conductivity $\alpha_{xy}$. We apply the finite-differences expression for Berry curvature\cite{FHS,FH, FENG20121849} to metallic systems.
One of the method's advantages is that $\sigma_{xy}$ and $\alpha_{xy}$ can be calculated without the construction of Wannier functions demanding the technical or empirical procedure. 
We implemented our proposed method in the OpenMX code\cite{openmx}, a first-principles calculation package, as a post-processing code. 
To confirm the consistency on our proposed method, we calculate $\sigma_{xy}$ and $\alpha_{xy}$ in two-dimensional ferromagnet FeCl$_2$ and three-dimensional ferromagnetic transition metals (bcc-Fe, hcp-Co, and fcc-Ni) as a practice.
Calculation results successfully reproduced the previous reported ones \cite{Rifky,YYao,Vanderbilt, CCLee, Guo_PRB_2011,Weischenberg_PRB_2013} obtained from the Kubo formula\cite{YYao} and Wannier representations \cite{Vanderbilt, Lopez}.
Our proposed method will open efficient thermoelectric materials design based on high-throughput first-principles screening.
The source code and input files are publicly available on GitHub\footnote{\url{https://github.com/hikaruri/OMXsigmaxy}}.
\section{Methods}

\begin{figure*}[tbp]
\includegraphics[width=2\columnwidth]{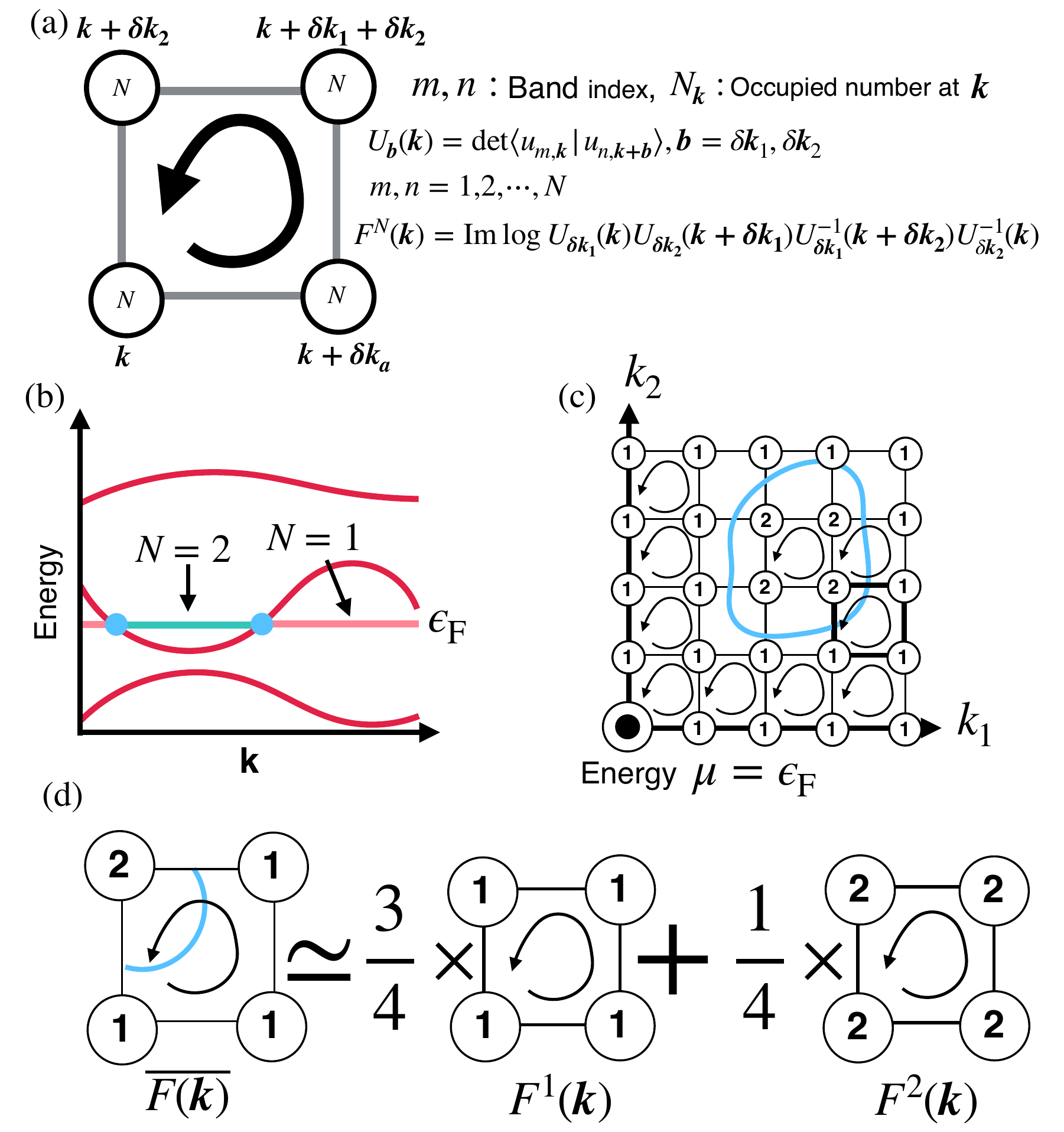}
\caption{
  (a) Computing $\sigma_{xy}$ in insulating system.
  (b) Fermi energy intersecting energy bands.
  (c) Carrying out contour integration on each plaquette. Except for the plaquette intersecting the Fermi surface, Berry curvature can be computed the same as insulating system because the occupation numbers $N$ on four vertexes are equal to one another.
(d) The schematic diagram of an approximation in computing the Berry curvature on the plaquette
intersecting the Fermi surface. Assuming occupation numbers on four vertexes are equal, one can take the average of the computed Berry curvature}
\label{RevFHS}
\end{figure*}

First, we explain a method of computing $\sigma_{xy}$ in an insulating system, which was proposed by Fukui-Hatsugai-Suzuki\cite{FHS, FH}.

We set the wave vector ${\bf k}=(k_{1},k_{2})$ at the lattice point on the two-dimensional Brillouin zone and defined its grid interval as $\delta {\bf k}_{1}$ and $\delta {\bf k}_{2}$ along the $k_{1}$ and $k_{2}$ directions, respectively.
The $N\times N$ overlap matrix on the Brillouin zone is defined as follows:
\begin{eqnarray}
  \label{OverlapMatrix}
(M_{{\bf k},{\bf k+\delta{\bf k}}})_{mn} = \langle u_m({\bf k})|u_n({\bf k}+\delta {\bf k})\rangle
\end{eqnarray}
and we defined U(1) link variable as follows:
\begin{eqnarray}
  \label{U1link}
  U_{\delta {\bf k}}({\bf k}) = \det M_{{\bf k},{\bf k+\delta{\bf k}}}.
\end{eqnarray}
The non-abelian Berry curvature\cite{FHS} on the Brillouin zone, the local Berry phase\cite{PhysRevB.47.1651}, can be computed using $U_{\delta {\bf k}}({\bf k})$ as follows:
\begin{eqnarray}
  \label{FHSBerryF}
F({\bf k}) =  {\rm Im} \ln 
U_{\delta {\bf k}_{1}}({\bf k})
U_{\delta {\bf k}_{2}}({\bf k}+\delta {\bf k}_{1}) \\ \nonumber
\times U_{\delta {\bf k}_{1}}({\bf k}+\delta {\bf k}_{2})^{-1}
U_{\delta {\bf k}_{2}}({\bf k})^{-1}.
\end{eqnarray}
The value of $F({\bf k})$ varies in the range of $-\pi\leq F({\bf k})<\pi$ because ${\rm Im}\ln$ is an operation to take the argument of a complex number.
To compute the Berry curvature $F$, we carried out the contour integration at four wave numbers on the vertices of a plaquette as shown in Fig. \ref{RevFHS}(a). The anomalous Hall conductivity, $\sigma_{xy}$ in the insulating system is computed by obtaining the Berry curvature on the Brillouin zone $F({\bf k})$
\begin{eqnarray}
\sigma_{xy}(\mu)= -\frac{e^2}{h}\frac{1}{2\pi}\sum_{\bf k}F({\bf k})
=-\frac{e^2}{h}C
\end{eqnarray}
For computing Eq.(\ref{FHSBerryF}), the matrix size of $U$ must be equal to that in another vertex, i.e., the all occupied number $N$ on vertices of a plaquette must be equal.
Therefore, this method can apply only 
in the insulating system which all occupation numbers $N$ on the vertices of a plaquette are equal.
\par
Next, we expanded the Fukui-Hatsugai-Suzuki method to metallic system. As shown in Fig. \ref{RevFHS}(b), we consider the case in which any band intersects the Fermi energy.
\par
(i) In the case of occupation numbers $N$ on four vertices of a plaquette are equal, we can compute Berry curvature similarly to the case of an insulating system. As shown in Fig. \ref{RevFHS}(c), we can compute $U$ on each plaquette and obtain Berry curvature $F$.
\par
(ii) In the case that even one occupation number $N$ is different from one of the plaquette, we approximate the Berry curvature $F$ by computing the average.
Figure \ref{RevFHS}(d) illustrates the approximation concept for determining the average $F$. 
For example, in the case where the occupation numbers on four vertexes are $N_{1}, N_{2}, N_{3},$ and $N_{4}$, we can obtain the four Berry curvatures $F_{1}({\bf k}), F_{2}({\bf k}), F_{3}({\bf k}),$ and $ F_{4}({\bf k})$ which are calculated assuming that the all occupation numbers on the four vertexes are $N_{1}, N_{2}, N_{3}$ and $N_{4}$. 
The approximated Berry curvature $\bar{F}$ on this plaquette is approximated by the following equation:
\begin{eqnarray}
\bar{F}({\bf k}) \simeq \frac{1}{4}\left(
  F_{1}({\bf k})
  +F_{2}({\bf k})
  +F_{3}({\bf k})
  +F_{4}({\bf k})
\right).\label{eq.Berry}
\end{eqnarray}
Through this approximation, we can compute $\bar{F}({\bf k})$ on all plaquette, and we can obtain $\sigma_{xy}$ as\par
\begin{eqnarray}
  \sigma_{xy}(\mu)
  =-\frac{e^2}{h}\sum_{{\bf k}}\bar{F}({\bf k}).
\end{eqnarray}
\par
To approximate $\alpha_{xy}$, we need to calculate the $\mu$ dependence of $\sigma_{xy}$.
If we set the occupation number $N$ corresponding to a chemical potential height, we can compute the chemical potential dependence of $\sigma_{xy}$ immediately
by computing the overlap matrices until the occupation number $N$ in Eq.(\ref{OverlapMatrix}).
If $\mu$ is changed, it is only necessary to recalculate the Eq.(\ref{U1link}) in each plaquette needs to be calculated.
Because the computational cost of Eq.(\ref{U1link}) is much smaller than that of Eq.(\ref{OverlapMatrix}), it is possible to calculate $\alpha_{xy}$ efficiently.
\par
Finally, in the case of a three-dimensional system, $\sigma_{xy}$ are defined on each ${\bf k}_{3}$.
Thus, $\sigma_{xy}$ in a bulk system is computed by the average along the ${\bf k}_{3}$ direction as follows:
\begin{eqnarray}
\sigma_{xy}(\mu)
=\frac{1}{N_3}\sum_{{\bf k}_{3}}\sigma_{xy}
(\mu,{\bf k}_{3}).
\end{eqnarray}
Here, $N_{3}$ is the mesh number along the ${\bf k}_{3}$ direction.
Through this, $\sigma_{xy}$ in a three-dimensional system can be computed by applying our method on each ${\bf k}_3$.
\section{Computational condition}
We conducted first-principles calculations based on the noncollinear density functional theory (DFT) based using the OpenMX code \cite{openmx}.
DFT calculations were performed through the exchange-correlation functional within the generalized gradient approximation\cite{PhysRevLett.77.3865} and norm-conserving pseudopotentials\cite{PhysRevLett.43.1494}. The wave functions were expanded by a linear combination of multiple pseudoatomic orbitals \cite{OzakiPRB67,OzakiKinoPRB69}.
The spin-orbit interaction was included by using total-angular-momentum-dependent pseudopotentials\cite{PhysRevB.64.073106}.
For FeCl$_2$, the cutoff energy for a charge density of 500 Ry, a ${\bf k}$-point sampling of $20\times20\times1$, and lattice constant of 3.475 {\AA} were used.
A set of pseudo atomic orbital basis functions was specified as Fe6.0S-$s2p3d3f1$ and Cl7.0-$s3p3d2$, where 6.0 and 7.0 are the cutoff radii (in bohrs) of each element, respectively. S stands for a soft pseudopotential, and the integers after $s, p, d,$ and $f$ indicates the radial multiplicity of each angular momentum component. These computational conditions are the same as those reported previously study\cite{Rifky}. For bcc-Fe, the cutoff energy for a charge density of 300 Ry, a ${\bf k}$-point sampling of $36\times36\times36$, and lattice constant of 2.87 {\AA} were used.
A set of pseudo atomic orbital basis functions was specified as Fe6.0S-$s3p3d3f1$. For details of hcp-Co and fcc-Ni, see Appendix \ref{appA}.
\section{Results and discussion}
\subsection{2D ferromagnetic materials}
\begin{figure*}[tbp]
\includegraphics[width=2.05\columnwidth]{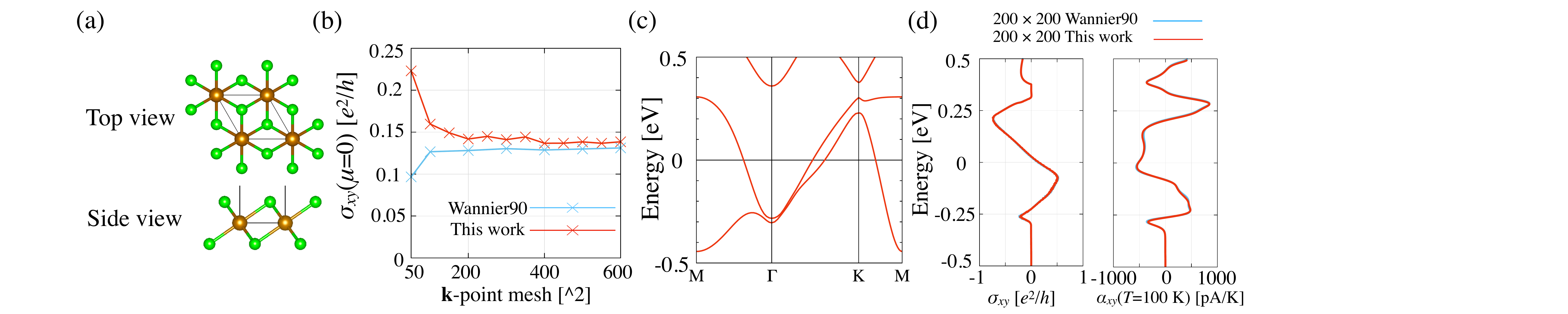}
\caption{
(a) Crystal structure of FeCl$_2$. (b) ${\bf k}$-mesh dependence of the $\sigma_{xy}$ at the Fermi energy.
(c) Band structure of FeCl$_2$. (d) Chemical potential dependence of the $\sigma_{xy}$ and $\alpha_{xy}$. 
Blue and Red solid line points correspond to the present calculation results and results obtained from the Wannier90, respectively.
}
\label{FeCl2}
\end{figure*}

Here, our method is applied for the two-dimensional ferromagnetic FeCl$_2$, which has a simple electronic structure and a large ANE\cite{Rifky}.
Figure \ref{FeCl2}(a) shows the schematic crystal structure of FeCl$_2$, where six Cl atoms connect each Fe atom. This material is known as a ferromagnetic two-dimensional material with half-metallicity\cite{Rifky, C7TC02664A, ZHENG2017184}. 
\par
Figure \ref{FeCl2}(b) shows the ${\bf k}$-mesh dependence of the $\sigma_{xy}$ at the Fermi energy.
We found that the $\sigma_{xy}$ converged to 0.14 $(e^2/h)$ with a $200\times 200$ ${\bf k}$-mesh
and its value converged within about 10 \% with a $100\times 100$ ${\bf k}$-mesh.
This convergence with a small number of ${\bf k}$-mesh is similar to the previous first-principles calculation\cite{PhysRevB.85.012405}. 
Compared to Wannier90's results, our results exhibit a very small difference less than 0.014$(e^2/h)$ where ${\bf k}$-mesh is greater than $200\times 200$ at the Fermi energy. One of the factors for this convergence is that the ratio of the plaquette number, with the approximation of Eq.(\ref{eq.Berry}), decreased from 12\% to 3\% while ${\bf k}$-mesh increased from $50\times 50$ to $200\times 200$. 
\par
Next, we confirm the chemical potential dependence of the $\sigma_{xy}$ and $\alpha_{xy}$.
Figure \ref{FeCl2}(c) shows the electronic band structure for FeCl$_2$. A simple band structure without degenerate points near the Fermi energy is noticeable in this case. Figure \ref{FeCl2}(d) shows the chemical potential dependence of the $\sigma_{xy}$ at 0 K and $\alpha_{xy}$ at 100 K. The ${\bf k}$-mesh of $200\times 200\times 1$ were used for both calculations.
Our calculation results (Blue solid line in Fig. \ref{FeCl2}(d)) reproduce completely those obtained from the Wannier representations. Therefore, we conclude that our method can reproduce the results obtained from the Wannier representations in a simple electronic structure case.

\subsection{3D ferromagnetic materials}
\begin{table}[tbp]
  \caption{Comparison of this work and previous ones for the $\sigma_{xy}$ in bcc-Fe.}  
  \begin{tabular}{cc} \hline\hline
    Refs. &  $\sigma_{xy}$ (S/cm) \\ \hline
    Y. Yao et al.\cite{YYao} & 751 \\
    X. Wang et al.\cite{Vanderbilt} & 756.76\\
    C-C. Lee et al.\cite{CCLee} & 750 \\
    Exp.\cite{Dheer} & 1032\\
    This work ($100\times100\times100$) & 770\\
    This work ($200\times200\times200$) & 788\\
    This work ($300\times300\times300$) & 782\\
    This work ($400\times400\times400$) & 786\\
    This work ($500\times500\times500$) & 785\\ 
    This work ($600\times600\times600$) & 790\\ 
    This work ($700\times700\times700$) & 788\\ 
    \hline \hline
  \end{tabular}
  \label{prevStudy}
\end{table}

Next, let us perform our method in a three-dimensional ferromagnetic system.
We calculated the $\sigma_{xy}$ and $\alpha_{xy}$ for bcc-Fe, hcp-Co, and fcc-Ni as a typical example.
Here, we focused on bcc-Fe [Fig. \ref{parameterfig}(a)\cite{Momma:db5098}] (for hcp-Co and fcc-Ni, see Appendix \ref{appA}). In table \ref{prevStudy}, we compared the $\sigma_{xy}$ for bcc-Fe at the Fermi energy for the present work with a previous study. We can see that our calculation results converge to approximately $\sigma_{xy} \simeq 750$ S/cm as similarly reported in previous theoretical calculation. 
\par
Figure \ref{parameterfig}(b) shows the ${\bf k}$-mesh dependence of the $\sigma_{xy}$ at the Fermi energy. 
The $\sigma_{xy}$ converged with at least the ${\bf k}$-mesh of $200\times 200 \times 200$ and its value converged within about 10 \% with $100\times 100 \times 100$ ${\bf k}$-mesh.
Our results differ in value from those reported by Wannier 90’s with less than 0.86 S/cm, where the ${\bf k}$-mesh is finer than $200\times 200 \times 200$ at the Fermi energy. We can conclude that our method could reproduce the $\sigma_{xy}$ at a specific chemical potential.
\par
Finally, we discuss the chemical potential dependence of the $\sigma_{xy}$ and $\alpha_{xy}$ for bcc-Fe. Figures \ref{parameterfig}(c) and (d) show the band structure of bcc-Fe and the chemical potential dependence of the $\sigma_{xy}$ at 0 K and $\alpha_{xy}$ at 100 K, respectively. Compared to the two-dimensional system, the correspondence of the $\sigma_{xy}$ between our method and the Wannier representation is slightly lowered at the specific energy region from -1.0 eV to -0.5 eV. This numerical error may originate from an entangled or degenerate electronic structure because the bcc-Fe system has many degenerate points stemming from point nodes\cite{PhysRevB.92.085138}. However, due to the smearing of the Fermi-Dirac distribution function, this inconsistency decreases as the temperature increases. In fact, the shown in Fig. \ref{parameterfig}, shows that our $\alpha_{xy}$ of bcc-Fe at 100 K results are almost consistent with those calculated by Wannier90. We safely conclude that our method has enough accuracy for evaluating $\alpha_{xy}$ in a finite temperature.
\begin{figure*}[tbp]
\includegraphics[width=2.05\columnwidth]{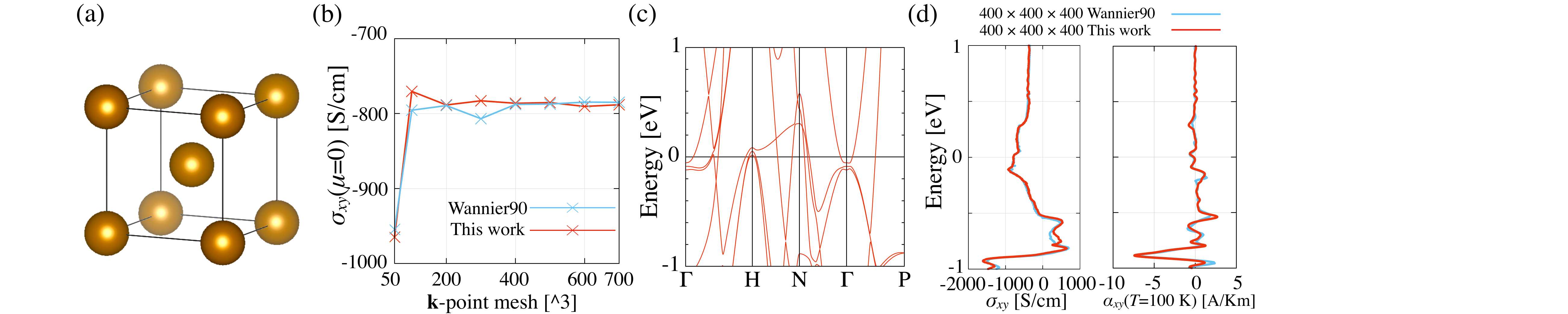}
\caption{
(a) Crystal structure of bcc-Fe. (b) ${\bf k}$-mesh dependence of the $\sigma_{xy}$ at the Fermi energy.
(c) Band structure of bcc-Fe. (d) Chemical potential dependence of the $\sigma_{xy}$ and $\alpha_{xy}$. 
Blue and Red solid line points correspond to the present calculation results and results obtained from the Wannier90, respectively.
}
\label{parameterfig}
\end{figure*}

\section{Conclusion}
In this study, we expanded the Fukui-Hatsugai-Suzuki method to a metallic system to improve the efficiency in calculation of $\sigma_{xy}$ and $\alpha_{xy}$ in magnetic materials. Calculating an average of the Berry curvature on the all ${\bf k}$-mesh plaquette, with respect to each vertex, makes it possible to estimate the $\sigma_{xy}$ in partially occupied cases. We also demonstrated the calculations of $\sigma_{xy}$ and $\alpha_{xy}$ by using this method in a typical two-dimensional ferromagnetic material FeCl$_2$ and three-dimensional magnetic transition metal bcc-Fe, hcp-Co, and fcc-Ni. The $\sigma_{xy}$ in FeCl$_2$ with a simple band structure completely reproduced the calculation results obtained from the Wannier representation and exhibited fast conversion with a rough ${\bf k}$-mesh. Whereas, in three-dimensional transition metal cases, the consistency is slightly dropped in a specific energy range because of a complicated band structure; however, we find a good agreement for anomalous Nernst conductivity at a finite temperature. The present study will give us a more efficient calculation method for the AHE and ANE without some technical and empirical procedures such as those constructing Wannier functions. High-throughput first-principles screening based on this method will be a useful tool for thermoelectric materials design.
\begin{acknowledgments}
  This work was partly supported by JSPS KAKENHI Grants Numbers JP20J13011, JP20K15115, JP20K22479, JP22K04862, JP22K14587. This work was partly supported by Japan Science and Technology Agency (JST) as part of SICORP, Grant Number  JPMJSC21E3.
  The computation was mainly carried out using the computer facilities at ISSP, The University of Tokyo and RIIT, Kyushu University.
  The computations in this research were partially performed using the Fujitsu PRIMERGY CX400M1/CX2550M5 (Oakbridge-CX) in the Information Technology Center, The University of Tokyo.
  Crystal structures were drawn by VESTA\cite{Momma:db5098}.
\end{acknowledgments}

\appendix
\section{Anomalous Hall conductivity for hcp-Co and fcc-Ni}\label{appA}

\begin{figure}[tbp]
\includegraphics[width=\columnwidth]{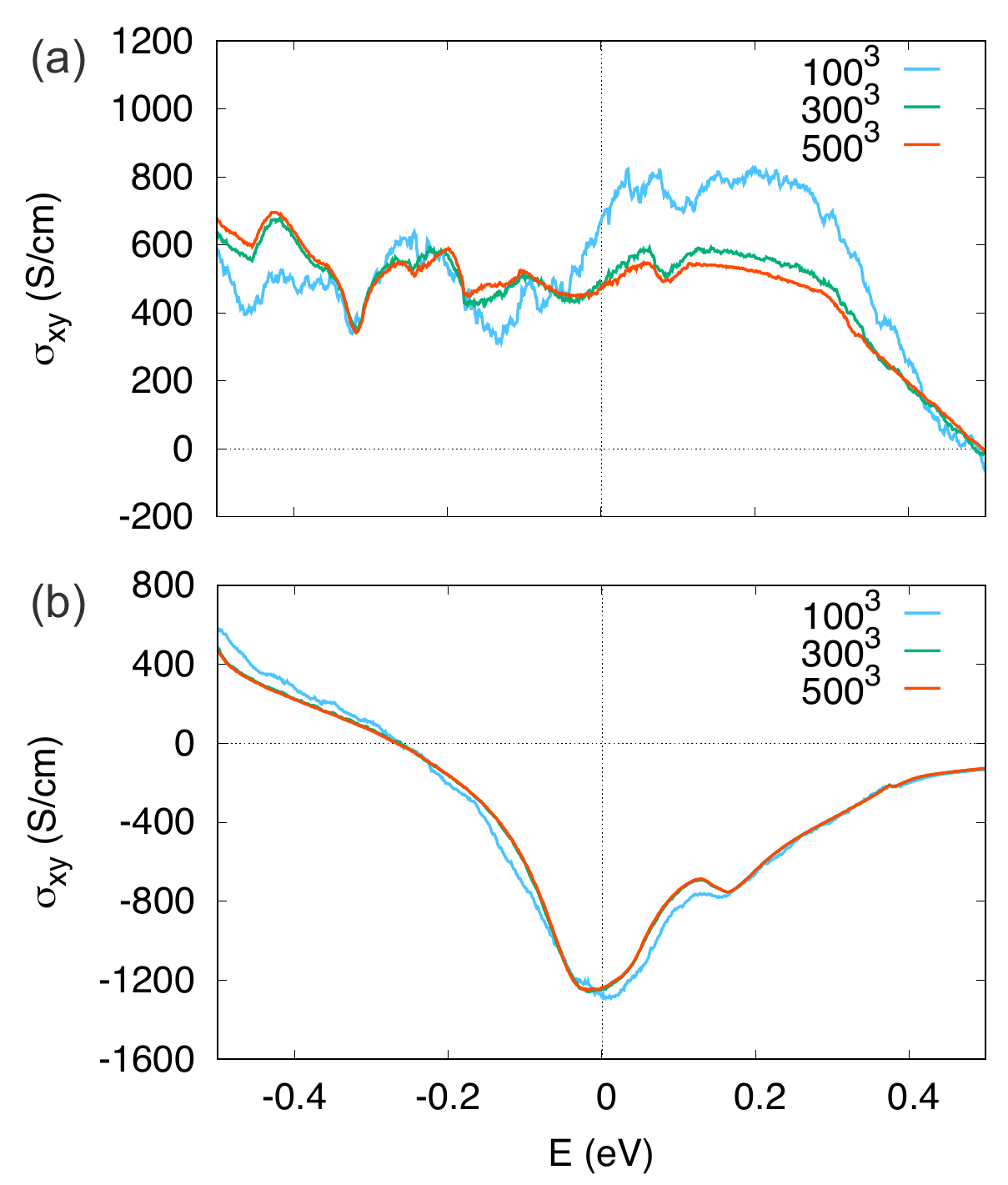}
\caption{
Chemical potential dependence of $\sigma_{xy}$ for (a) hcp-Co, and (b) fcc-Ni.
Blue, green and red solid lines correspond to the $k$-mesh of $100\times 100 \times 100$, $300\times 300 \times 300$ and $500\times 500 \times 500$, respectively.
}
\label{Co-Ni}
\end{figure}

We also calculated the $\sigma_{xy}$ for typical transition metal ferromagnetic materials of hcp-Co and fcc-Ni.
For fcc-Ni, the cutoff energy for a charge density of 300 Ry, a ${\bf k}$-point sampling of $32 \times 32 \times 32$, and lattice constant of $a=3.56$ {\AA} were used.
A set of pseudo atomic orbital basis functions was specified as  Ni6.0-s3p2d2f1.
For hcp-Co, the cutoff energy for a charge density of 300 Ry, a ${\bf k}$-point sampling of $24 \times 24 \times 18$, and lattice constant of $a=2.50$ {\AA} and $c=4.07$ {\AA} were used.
A set of pseudo atomic orbital basis functions was specified as Co6.0-s3p2d2f1. 
Figure \ref{Co-Ni} shows the chemical potential dependence of the $\sigma_{xy}$ for hcp-Co and fcc-Ni. We found that the ${\bf k}$-point mesh of $300\times 300 \times 300$ converged enough.
Our calculation results well reproduce the previous studies one. \cite{Guo_PRB_2011, Weischenberg_PRB_2013}

\bibliographystyle{apsrev4-1}
\bibliography{sigma_ref}

\begin{thebibliography}{67}%
\makeatletter
\providecommand \@ifxundefined [1]{%
 \@ifx{#1\undefined}
}%
\providecommand \@ifnum [1]{%
 \ifnum #1\expandafter \@firstoftwo
 \else \expandafter \@secondoftwo
 \fi
}%
\providecommand \@ifx [1]{%
 \ifx #1\expandafter \@firstoftwo
 \else \expandafter \@secondoftwo
 \fi
}%
\providecommand \natexlab [1]{#1}%
\providecommand \enquote  [1]{``#1''}%
\providecommand \bibnamefont  [1]{#1}%
\providecommand \bibfnamefont [1]{#1}%
\providecommand \citenamefont [1]{#1}%
\providecommand \href@noop [0]{\@secondoftwo}%
\providecommand \href [0]{\begingroup \@sanitize@url \@href}%
\providecommand \@href[1]{\@@startlink{#1}\@@href}%
\providecommand \@@href[1]{\endgroup#1\@@endlink}%
\providecommand \@sanitize@url [0]{\catcode `\\12\catcode `\$12\catcode
  `\&12\catcode `\#12\catcode `\^12\catcode `\_12\catcode `\%12\relax}%
\providecommand \@@startlink[1]{}%
\providecommand \@@endlink[0]{}%
\providecommand \url  [0]{\begingroup\@sanitize@url \@url }%
\providecommand \@url [1]{\endgroup\@href {#1}{\urlprefix }}%
\providecommand \urlprefix  [0]{URL }%
\providecommand \Eprint [0]{\href }%
\providecommand \doibase [0]{http://dx.doi.org/}%
\providecommand \selectlanguage [0]{\@gobble}%
\providecommand \bibinfo  [0]{\@secondoftwo}%
\providecommand \bibfield  [0]{\@secondoftwo}%
\providecommand \translation [1]{[#1]}%
\providecommand \BibitemOpen [0]{}%
\providecommand \bibitemStop [0]{}%
\providecommand \bibitemNoStop [0]{.\EOS\space}%
\providecommand \EOS [0]{\spacefactor3000\relax}%
\providecommand \BibitemShut  [1]{\csname bibitem#1\endcsname}%
\let\auto@bib@innerbib\@empty
\bibitem [{\citenamefont {Hall}(1880)}]{Hall}%
  \BibitemOpen
  \bibfield  {author} {\bibinfo {author} {\bibfnamefont {E.~H.}\ \bibnamefont
  {Hall}},\ }\href {\doibase 10.1088/1478-7814/4/1/335} {\bibfield  {journal}
  {\bibinfo  {journal} {Proceedings of the Physical Society of London}\
  }\textbf {\bibinfo {volume} {4}},\ \bibinfo {pages} {325} (\bibinfo {year}
  {1880})}\BibitemShut {NoStop}%
\bibitem [{\citenamefont {Nagaosa}\ \emph {et~al.}(2010)\citenamefont
  {Nagaosa}, \citenamefont {Sinova}, \citenamefont {Onoda}, \citenamefont
  {MacDonald},\ and\ \citenamefont {Ong}}]{Nagaosa}%
  \BibitemOpen
  \bibfield  {author} {\bibinfo {author} {\bibfnamefont {N.}~\bibnamefont
  {Nagaosa}}, \bibinfo {author} {\bibfnamefont {J.}~\bibnamefont {Sinova}},
  \bibinfo {author} {\bibfnamefont {S.}~\bibnamefont {Onoda}}, \bibinfo
  {author} {\bibfnamefont {A.~H.}\ \bibnamefont {MacDonald}}, \ and\ \bibinfo
  {author} {\bibfnamefont {N.~P.}\ \bibnamefont {Ong}},\ }\href {\doibase
  10.1103/RevModPhys.82.1539} {\bibfield  {journal} {\bibinfo  {journal} {Rev.
  Mod. Phys.}\ }\textbf {\bibinfo {volume} {82}},\ \bibinfo {pages} {1539}
  (\bibinfo {year} {2010})}\BibitemShut {NoStop}%
\bibitem [{\citenamefont {Smit}(1958)}]{JSmit}%
  \BibitemOpen
  \bibfield  {author} {\bibinfo {author} {\bibfnamefont {J.}~\bibnamefont
  {Smit}},\ }\href {\doibase https://doi.org/10.1016/S0031-8914(58)93541-9}
  {\bibfield  {journal} {\bibinfo  {journal} {Physica}\ }\textbf {\bibinfo
  {volume} {24}},\ \bibinfo {pages} {39} (\bibinfo {year} {1958})}\BibitemShut
  {NoStop}%
\bibitem [{\citenamefont {Berger}(1970)}]{LBerger}%
  \BibitemOpen
  \bibfield  {author} {\bibinfo {author} {\bibfnamefont {L.}~\bibnamefont
  {Berger}},\ }\href {\doibase 10.1103/PhysRevB.2.4559} {\bibfield  {journal}
  {\bibinfo  {journal} {Phys. Rev. B}\ }\textbf {\bibinfo {volume} {2}},\
  \bibinfo {pages} {4559} (\bibinfo {year} {1970})}\BibitemShut {NoStop}%
\bibitem [{\citenamefont {Berry}(1984)}]{Berry}%
  \BibitemOpen
  \bibfield  {author} {\bibinfo {author} {\bibfnamefont {M.~V.}\ \bibnamefont
  {Berry}},\ }\href {\doibase 10.1098/rspa.1984.0023} {\bibfield  {journal}
  {\bibinfo  {journal} {Proceedings of the Royal Society of London. A.
  Mathematical and Physical Sciences}\ }\textbf {\bibinfo {volume} {392}},\
  \bibinfo {pages} {45} (\bibinfo {year} {1984})}\BibitemShut {NoStop}%
\bibitem [{\citenamefont {Kohmoto}(1985)}]{Kohmoto}%
  \BibitemOpen
  \bibfield  {author} {\bibinfo {author} {\bibfnamefont {M.}~\bibnamefont
  {Kohmoto}},\ }\href {\doibase https://doi.org/10.1016/0003-4916(85)90148-4}
  {\bibfield  {journal} {\bibinfo  {journal} {Annals of Physics}\ }\textbf
  {\bibinfo {volume} {160}},\ \bibinfo {pages} {343} (\bibinfo {year}
  {1985})}\BibitemShut {NoStop}%
\bibitem [{\citenamefont {Xiao}\ \emph {et~al.}(2010)\citenamefont {Xiao},
  \citenamefont {Chang},\ and\ \citenamefont {Niu}}]{DXiao_RevModPhy_2010}%
  \BibitemOpen
  \bibfield  {author} {\bibinfo {author} {\bibfnamefont {D.}~\bibnamefont
  {Xiao}}, \bibinfo {author} {\bibfnamefont {M.-C.}\ \bibnamefont {Chang}}, \
  and\ \bibinfo {author} {\bibfnamefont {Q.}~\bibnamefont {Niu}},\ }\href
  {\doibase 10.1103/RevModPhys.82.1959} {\bibfield  {journal} {\bibinfo
  {journal} {Rev. Mod. Phys.}\ }\textbf {\bibinfo {volume} {82}},\ \bibinfo
  {pages} {1959} (\bibinfo {year} {2010})}\BibitemShut {NoStop}%
\bibitem [{\citenamefont {Ohgushi}\ \emph {et~al.}(2000)\citenamefont
  {Ohgushi}, \citenamefont {Murakami},\ and\ \citenamefont
  {Nagaosa}}]{Ohgushi}%
  \BibitemOpen
  \bibfield  {author} {\bibinfo {author} {\bibfnamefont {K.}~\bibnamefont
  {Ohgushi}}, \bibinfo {author} {\bibfnamefont {S.}~\bibnamefont {Murakami}}, \
  and\ \bibinfo {author} {\bibfnamefont {N.}~\bibnamefont {Nagaosa}},\ }\href
  {\doibase 10.1103/PhysRevB.62.R6065} {\bibfield  {journal} {\bibinfo
  {journal} {Phys. Rev. B}\ }\textbf {\bibinfo {volume} {62}},\ \bibinfo
  {pages} {R6065} (\bibinfo {year} {2000})}\BibitemShut {NoStop}%
\bibitem [{\citenamefont {Nakatsuji}\ \emph {et~al.}(2015)\citenamefont
  {Nakatsuji}, \citenamefont {Kiyohara},\ and\ \citenamefont
  {Higo}}]{Nakatsuji}%
  \BibitemOpen
  \bibfield  {author} {\bibinfo {author} {\bibfnamefont {S.}~\bibnamefont
  {Nakatsuji}}, \bibinfo {author} {\bibfnamefont {N.}~\bibnamefont {Kiyohara}},
  \ and\ \bibinfo {author} {\bibfnamefont {T.}~\bibnamefont {Higo}},\ }\href
  {\doibase 10.1038/nature15723} {\bibfield  {journal} {\bibinfo  {journal}
  {Nature}\ }\textbf {\bibinfo {volume} {527}},\ \bibinfo {pages} {212}
  (\bibinfo {year} {2015})}\BibitemShut {NoStop}%
\bibitem [{\citenamefont {Thouless}\ \emph {et~al.}(1982)\citenamefont
  {Thouless}, \citenamefont {Kohmoto}, \citenamefont {Nightingale},\ and\
  \citenamefont {den Nijs}}]{TKNN}%
  \BibitemOpen
  \bibfield  {author} {\bibinfo {author} {\bibfnamefont {D.~J.}\ \bibnamefont
  {Thouless}}, \bibinfo {author} {\bibfnamefont {M.}~\bibnamefont {Kohmoto}},
  \bibinfo {author} {\bibfnamefont {M.~P.}\ \bibnamefont {Nightingale}}, \ and\
  \bibinfo {author} {\bibfnamefont {M.}~\bibnamefont {den Nijs}},\ }\href
  {\doibase 10.1103/PhysRevLett.49.405} {\bibfield  {journal} {\bibinfo
  {journal} {Phys. Rev. Lett.}\ }\textbf {\bibinfo {volume} {49}},\ \bibinfo
  {pages} {405} (\bibinfo {year} {1982})}\BibitemShut {NoStop}%
\bibitem [{\citenamefont {Chang}\ \emph {et~al.}(2013)\citenamefont {Chang},
  \citenamefont {Zhang}, \citenamefont {Feng}, \citenamefont {Shen},
  \citenamefont {Zhang}, \citenamefont {Guo}, \citenamefont {Li}, \citenamefont
  {Ou}, \citenamefont {Wei}, \citenamefont {Wang}, \citenamefont {Ji},
  \citenamefont {Feng}, \citenamefont {Ji}, \citenamefont {Chen}, \citenamefont
  {Jia}, \citenamefont {Dai}, \citenamefont {Fang}, \citenamefont {Zhang},
  \citenamefont {He}, \citenamefont {Wang}, \citenamefont {Lu}, \citenamefont
  {Ma},\ and\ \citenamefont {Xue}}]{Science}%
  \BibitemOpen
  \bibfield  {author} {\bibinfo {author} {\bibfnamefont {C.-Z.}\ \bibnamefont
  {Chang}}, \bibinfo {author} {\bibfnamefont {J.}~\bibnamefont {Zhang}},
  \bibinfo {author} {\bibfnamefont {X.}~\bibnamefont {Feng}}, \bibinfo {author}
  {\bibfnamefont {J.}~\bibnamefont {Shen}}, \bibinfo {author} {\bibfnamefont
  {Z.}~\bibnamefont {Zhang}}, \bibinfo {author} {\bibfnamefont
  {M.}~\bibnamefont {Guo}}, \bibinfo {author} {\bibfnamefont {K.}~\bibnamefont
  {Li}}, \bibinfo {author} {\bibfnamefont {Y.}~\bibnamefont {Ou}}, \bibinfo
  {author} {\bibfnamefont {P.}~\bibnamefont {Wei}}, \bibinfo {author}
  {\bibfnamefont {L.-L.}\ \bibnamefont {Wang}}, \bibinfo {author}
  {\bibfnamefont {Z.-Q.}\ \bibnamefont {Ji}}, \bibinfo {author} {\bibfnamefont
  {Y.}~\bibnamefont {Feng}}, \bibinfo {author} {\bibfnamefont {S.}~\bibnamefont
  {Ji}}, \bibinfo {author} {\bibfnamefont {X.}~\bibnamefont {Chen}}, \bibinfo
  {author} {\bibfnamefont {J.}~\bibnamefont {Jia}}, \bibinfo {author}
  {\bibfnamefont {X.}~\bibnamefont {Dai}}, \bibinfo {author} {\bibfnamefont
  {Z.}~\bibnamefont {Fang}}, \bibinfo {author} {\bibfnamefont {S.-C.}\
  \bibnamefont {Zhang}}, \bibinfo {author} {\bibfnamefont {K.}~\bibnamefont
  {He}}, \bibinfo {author} {\bibfnamefont {Y.}~\bibnamefont {Wang}}, \bibinfo
  {author} {\bibfnamefont {L.}~\bibnamefont {Lu}}, \bibinfo {author}
  {\bibfnamefont {X.-C.}\ \bibnamefont {Ma}}, \ and\ \bibinfo {author}
  {\bibfnamefont {Q.-K.}\ \bibnamefont {Xue}},\ }\href {\doibase
  10.1126/science.1234414} {\bibfield  {journal} {\bibinfo  {journal}
  {Science}\ }\textbf {\bibinfo {volume} {340}},\ \bibinfo {pages} {167}
  (\bibinfo {year} {2013})}\BibitemShut {NoStop}%
\bibitem [{\citenamefont {Xiao}\ \emph {et~al.}(2006)\citenamefont {Xiao},
  \citenamefont {Yao}, \citenamefont {Fang},\ and\ \citenamefont
  {Niu}}]{DXiao}%
  \BibitemOpen
  \bibfield  {author} {\bibinfo {author} {\bibfnamefont {D.}~\bibnamefont
  {Xiao}}, \bibinfo {author} {\bibfnamefont {Y.}~\bibnamefont {Yao}}, \bibinfo
  {author} {\bibfnamefont {Z.}~\bibnamefont {Fang}}, \ and\ \bibinfo {author}
  {\bibfnamefont {Q.}~\bibnamefont {Niu}},\ }\href {\doibase
  10.1103/PhysRevLett.97.026603} {\bibfield  {journal} {\bibinfo  {journal}
  {Phys. Rev. Lett.}\ }\textbf {\bibinfo {volume} {97}},\ \bibinfo {pages}
  {026603} (\bibinfo {year} {2006})}\BibitemShut {NoStop}%
\bibitem [{\citenamefont {Sakuraba}(2016)}]{Sakuraba}%
  \BibitemOpen
  \bibfield  {author} {\bibinfo {author} {\bibfnamefont {Y.}~\bibnamefont
  {Sakuraba}},\ }\href {\doibase
  https://doi.org/10.1016/j.scriptamat.2015.04.034} {\bibfield  {journal}
  {\bibinfo  {journal} {Scripta Mater.}\ }\textbf {\bibinfo {volume} {111}},\
  \bibinfo {pages} {29 } (\bibinfo {year} {2016})}\BibitemShut {NoStop}%
\bibitem [{\citenamefont {Mizuguchi}\ and\ \citenamefont
  {Nakatsuji}(2019)}]{Mizuguchi_SciTechAdvMat_2019}%
  \BibitemOpen
  \bibfield  {author} {\bibinfo {author} {\bibfnamefont {M.}~\bibnamefont
  {Mizuguchi}}\ and\ \bibinfo {author} {\bibfnamefont {S.}~\bibnamefont
  {Nakatsuji}},\ }\href {\doibase 10.1080/14686996.2019.1585143} {\bibfield
  {journal} {\bibinfo  {journal} {Sci. Tech. Adv. Mater}\ }\textbf {\bibinfo
  {volume} {20}},\ \bibinfo {pages} {262} (\bibinfo {year} {2019})}\BibitemShut
  {NoStop}%
\bibitem [{\citenamefont {Lee}\ \emph {et~al.}(2004)\citenamefont {Lee},
  \citenamefont {Watauchi}, \citenamefont {Miller}, \citenamefont {Cava},\ and\
  \citenamefont {Ong}}]{Lee_PRL_2004}%
  \BibitemOpen
  \bibfield  {author} {\bibinfo {author} {\bibfnamefont {W.-L.}\ \bibnamefont
  {Lee}}, \bibinfo {author} {\bibfnamefont {S.}~\bibnamefont {Watauchi}},
  \bibinfo {author} {\bibfnamefont {V.~L.}\ \bibnamefont {Miller}}, \bibinfo
  {author} {\bibfnamefont {R.~J.}\ \bibnamefont {Cava}}, \ and\ \bibinfo
  {author} {\bibfnamefont {N.~P.}\ \bibnamefont {Ong}},\ }\href {\doibase
  10.1103/PhysRevLett.93.226601} {\bibfield  {journal} {\bibinfo  {journal}
  {Phys. Rev. Lett.}\ }\textbf {\bibinfo {volume} {93}},\ \bibinfo {pages}
  {226601} (\bibinfo {year} {2004})}\BibitemShut {NoStop}%
\bibitem [{\citenamefont {Miyasato}\ \emph {et~al.}(2007)\citenamefont
  {Miyasato}, \citenamefont {Abe}, \citenamefont {Fujii}, \citenamefont
  {Asamitsu}, \citenamefont {Onoda}, \citenamefont {Onose}, \citenamefont
  {Nagaosa},\ and\ \citenamefont {Tokura}}]{Miyasato_PRL_2007}%
  \BibitemOpen
  \bibfield  {author} {\bibinfo {author} {\bibfnamefont {T.}~\bibnamefont
  {Miyasato}}, \bibinfo {author} {\bibfnamefont {N.}~\bibnamefont {Abe}},
  \bibinfo {author} {\bibfnamefont {T.}~\bibnamefont {Fujii}}, \bibinfo
  {author} {\bibfnamefont {A.}~\bibnamefont {Asamitsu}}, \bibinfo {author}
  {\bibfnamefont {S.}~\bibnamefont {Onoda}}, \bibinfo {author} {\bibfnamefont
  {Y.}~\bibnamefont {Onose}}, \bibinfo {author} {\bibfnamefont
  {N.}~\bibnamefont {Nagaosa}}, \ and\ \bibinfo {author} {\bibfnamefont
  {Y.}~\bibnamefont {Tokura}},\ }\href {\doibase 10.1103/PhysRevLett.99.086602}
  {\bibfield  {journal} {\bibinfo  {journal} {Phys. Rev. Lett.}\ }\textbf
  {\bibinfo {volume} {99}},\ \bibinfo {pages} {086602} (\bibinfo {year}
  {2007})}\BibitemShut {NoStop}%
\bibitem [{\citenamefont {Pu}\ \emph {et~al.}(2008)\citenamefont {Pu},
  \citenamefont {Chiba}, \citenamefont {Matsukura}, \citenamefont {Ohno},\ and\
  \citenamefont {Shi}}]{Yong_PRL_2008}%
  \BibitemOpen
  \bibfield  {author} {\bibinfo {author} {\bibfnamefont {Y.}~\bibnamefont
  {Pu}}, \bibinfo {author} {\bibfnamefont {D.}~\bibnamefont {Chiba}}, \bibinfo
  {author} {\bibfnamefont {F.}~\bibnamefont {Matsukura}}, \bibinfo {author}
  {\bibfnamefont {H.}~\bibnamefont {Ohno}}, \ and\ \bibinfo {author}
  {\bibfnamefont {J.}~\bibnamefont {Shi}},\ }\href {\doibase
  10.1103/PhysRevLett.101.117208} {\bibfield  {journal} {\bibinfo  {journal}
  {Phys. Rev. Lett.}\ }\textbf {\bibinfo {volume} {101}},\ \bibinfo {pages}
  {117208} (\bibinfo {year} {2008})}\BibitemShut {NoStop}%
\bibitem [{\citenamefont {Sakuraba}\ \emph {et~al.}(2013)\citenamefont
  {Sakuraba}, \citenamefont {Hasegawa}, \citenamefont {Mizuguchi},
  \citenamefont {Kubota}, \citenamefont {Mizukami}, \citenamefont {Miyazaki},\
  and\ \citenamefont {Takanashi}}]{Sakuraba_APEX_2013}%
  \BibitemOpen
  \bibfield  {author} {\bibinfo {author} {\bibfnamefont {Y.}~\bibnamefont
  {Sakuraba}}, \bibinfo {author} {\bibfnamefont {K.}~\bibnamefont {Hasegawa}},
  \bibinfo {author} {\bibfnamefont {M.}~\bibnamefont {Mizuguchi}}, \bibinfo
  {author} {\bibfnamefont {T.}~\bibnamefont {Kubota}}, \bibinfo {author}
  {\bibfnamefont {S.}~\bibnamefont {Mizukami}}, \bibinfo {author}
  {\bibfnamefont {T.}~\bibnamefont {Miyazaki}}, \ and\ \bibinfo {author}
  {\bibfnamefont {K.}~\bibnamefont {Takanashi}},\ }\href {\doibase
  10.7567/apex.6.033003} {\bibfield  {journal} {\bibinfo  {journal} {Appl.
  Phys. Express}\ }\textbf {\bibinfo {volume} {6}},\ \bibinfo {pages} {033003}
  (\bibinfo {year} {2013})}\BibitemShut {NoStop}%
\bibitem [{\citenamefont {Hasegawa}\ \emph {et~al.}(2015)\citenamefont
  {Hasegawa}, \citenamefont {Mizuguchi}, \citenamefont {Sakuraba},
  \citenamefont {Kamada}, \citenamefont {Kojima}, \citenamefont {Kubota},
  \citenamefont {Mizukami}, \citenamefont {Miyazaki},\ and\ \citenamefont
  {Takanashi}}]{doi:10.1063/1.4922901}%
  \BibitemOpen
  \bibfield  {author} {\bibinfo {author} {\bibfnamefont {K.}~\bibnamefont
  {Hasegawa}}, \bibinfo {author} {\bibfnamefont {M.}~\bibnamefont {Mizuguchi}},
  \bibinfo {author} {\bibfnamefont {Y.}~\bibnamefont {Sakuraba}}, \bibinfo
  {author} {\bibfnamefont {T.}~\bibnamefont {Kamada}}, \bibinfo {author}
  {\bibfnamefont {T.}~\bibnamefont {Kojima}}, \bibinfo {author} {\bibfnamefont
  {T.}~\bibnamefont {Kubota}}, \bibinfo {author} {\bibfnamefont
  {S.}~\bibnamefont {Mizukami}}, \bibinfo {author} {\bibfnamefont
  {T.}~\bibnamefont {Miyazaki}}, \ and\ \bibinfo {author} {\bibfnamefont
  {K.}~\bibnamefont {Takanashi}},\ }\href {https://doi.org/10.1063/1.4922901}
  {\bibfield  {journal} {\bibinfo  {journal} {Appl. Phys. Lett.}\ }\textbf
  {\bibinfo {volume} {106}},\ \bibinfo {pages} {252405} (\bibinfo {year}
  {2015})}\BibitemShut {NoStop}%
\bibitem [{\citenamefont {Mizuta}\ and\ \citenamefont
  {Ishii}(2016)}]{YMizuta_H}%
  \BibitemOpen
  \bibfield  {author} {\bibinfo {author} {\bibfnamefont {Y.~P.}\ \bibnamefont
  {Mizuta}}\ and\ \bibinfo {author} {\bibfnamefont {F.}~\bibnamefont {Ishii}},\
  }\href {\doibase 10.1038/srep28076} {\bibfield  {journal} {\bibinfo
  {journal} {Sci. Rep.}\ }\textbf {\bibinfo {volume} {6}},\ \bibinfo {pages}
  {28076} (\bibinfo {year} {2016})}\BibitemShut {NoStop}%
\bibitem [{\citenamefont {Ikhlas}\ \emph {et~al.}(2017)\citenamefont {Ikhlas},
  \citenamefont {Tomita}, \citenamefont {Koretsune}, \citenamefont {Suzuki},
  \citenamefont {Nishio-Hamane}, \citenamefont {Arita}, \citenamefont {Otani},\
  and\ \citenamefont {Nakatsuji}}]{Mn3Sn}%
  \BibitemOpen
  \bibfield  {author} {\bibinfo {author} {\bibfnamefont {M.}~\bibnamefont
  {Ikhlas}}, \bibinfo {author} {\bibfnamefont {T.}~\bibnamefont {Tomita}},
  \bibinfo {author} {\bibfnamefont {T.}~\bibnamefont {Koretsune}}, \bibinfo
  {author} {\bibfnamefont {M.-T.}\ \bibnamefont {Suzuki}}, \bibinfo {author}
  {\bibfnamefont {D.}~\bibnamefont {Nishio-Hamane}}, \bibinfo {author}
  {\bibfnamefont {R.}~\bibnamefont {Arita}}, \bibinfo {author} {\bibfnamefont
  {Y.}~\bibnamefont {Otani}}, \ and\ \bibinfo {author} {\bibfnamefont
  {S.}~\bibnamefont {Nakatsuji}},\ }\href {\doibase 10.1038/nphys4181}
  {\bibfield  {journal} {\bibinfo  {journal} {Nat. Phys.}\ }\textbf {\bibinfo
  {volume} {13}},\ \bibinfo {pages} {1085} (\bibinfo {year}
  {2017})}\BibitemShut {NoStop}%
\bibitem [{\citenamefont {Guo}\ and\ \citenamefont
  {Wang}(2017)}]{PhysRevB.96.224415}%
  \BibitemOpen
  \bibfield  {author} {\bibinfo {author} {\bibfnamefont {G.-Y.}\ \bibnamefont
  {Guo}}\ and\ \bibinfo {author} {\bibfnamefont {T.-C.}\ \bibnamefont {Wang}},\
  }\href {\doibase 10.1103/PhysRevB.96.224415} {\bibfield  {journal} {\bibinfo
  {journal} {Phys. Rev. B}\ }\textbf {\bibinfo {volume} {96}},\ \bibinfo
  {pages} {224415} (\bibinfo {year} {2017})}\BibitemShut {NoStop}%
\bibitem [{\citenamefont {Mizuta}\ \emph {et~al.}(2018)\citenamefont {Mizuta},
  \citenamefont {Sawahata},\ and\ \citenamefont {Ishii}}]{YMizuta_EuO}%
  \BibitemOpen
  \bibfield  {author} {\bibinfo {author} {\bibfnamefont {Y.~P.}\ \bibnamefont
  {Mizuta}}, \bibinfo {author} {\bibfnamefont {H.}~\bibnamefont {Sawahata}}, \
  and\ \bibinfo {author} {\bibfnamefont {F.}~\bibnamefont {Ishii}},\ }\href
  {\doibase 10.1103/PhysRevB.98.205125} {\bibfield  {journal} {\bibinfo
  {journal} {Phys. Rev. B}\ }\textbf {\bibinfo {volume} {98}},\ \bibinfo
  {pages} {205125} (\bibinfo {year} {2018})}\BibitemShut {NoStop}%
\bibitem [{\citenamefont {Minami}\ \emph {et~al.}(2018)\citenamefont {Minami},
  \citenamefont {Ishii}, \citenamefont {Mizuta},\ and\ \citenamefont
  {Saito}}]{Minami_APL_2018}%
  \BibitemOpen
  \bibfield  {author} {\bibinfo {author} {\bibfnamefont {S.}~\bibnamefont
  {Minami}}, \bibinfo {author} {\bibfnamefont {F.}~\bibnamefont {Ishii}},
  \bibinfo {author} {\bibfnamefont {Y.~P.}\ \bibnamefont {Mizuta}}, \ and\
  \bibinfo {author} {\bibfnamefont {M.}~\bibnamefont {Saito}},\ }\href
  {\doibase 10.1063/1.5029907} {\bibfield  {journal} {\bibinfo  {journal}
  {Appl. Phys. Lett.}\ }\textbf {\bibinfo {volume} {113}},\ \bibinfo {pages}
  {032403} (\bibinfo {year} {2018})}\BibitemShut {NoStop}%
\bibitem [{\citenamefont {Sakai}\ \emph {et~al.}(2018)\citenamefont {Sakai},
  \citenamefont {Mizuta}, \citenamefont {Nugroho}, \citenamefont {Sihombing},
  \citenamefont {Koretsune}, \citenamefont {Suzuki}, \citenamefont {Takemori},
  \citenamefont {Ishii}, \citenamefont {Nishio-Hamane}, \citenamefont {Arita},
  \citenamefont {Goswami},\ and\ \citenamefont {Nakatsuji}}]{Sakai}%
  \BibitemOpen
  \bibfield  {author} {\bibinfo {author} {\bibfnamefont {A.}~\bibnamefont
  {Sakai}}, \bibinfo {author} {\bibfnamefont {Y.}~\bibnamefont {Mizuta}},
  \bibinfo {author} {\bibfnamefont {A.}~\bibnamefont {Nugroho}}, \bibinfo
  {author} {\bibfnamefont {R.}~\bibnamefont {Sihombing}}, \bibinfo {author}
  {\bibfnamefont {T.}~\bibnamefont {Koretsune}}, \bibinfo {author}
  {\bibfnamefont {M.}~\bibnamefont {Suzuki}}, \bibinfo {author} {\bibfnamefont
  {N.}~\bibnamefont {Takemori}}, \bibinfo {author} {\bibfnamefont
  {R.}~\bibnamefont {Ishii}}, \bibinfo {author} {\bibfnamefont
  {D.}~\bibnamefont {Nishio-Hamane}}, \bibinfo {author} {\bibfnamefont
  {R.}~\bibnamefont {Arita}}, \bibinfo {author} {\bibfnamefont
  {P.}~\bibnamefont {Goswami}}, \ and\ \bibinfo {author} {\bibfnamefont
  {S.}~\bibnamefont {Nakatsuji}},\ }\href {\doibase 10.1038/s41567-018-0225-6}
  {\bibfield  {journal} {\bibinfo  {journal} {Nat. Phys.}\ }\textbf {\bibinfo
  {volume} {14}},\ \bibinfo {pages} {1119} (\bibinfo {year}
  {2018})}\BibitemShut {NoStop}%
\bibitem [{\citenamefont {Liu}\ \emph {et~al.}(2018)\citenamefont {Liu},
  \citenamefont {Sun}, \citenamefont {Kumar}, \citenamefont {Muechler},
  \citenamefont {Sun}, \citenamefont {Jiao}, \citenamefont {Yang},
  \citenamefont {Liu}, \citenamefont {Liang}, \citenamefont {Xu}, \citenamefont
  {Kroder}, \citenamefont {S{\"u}{\ss}}, \citenamefont {Borrmann},
  \citenamefont {Shekhar}, \citenamefont {Wang}, \citenamefont {Xi},
  \citenamefont {Wang}, \citenamefont {Schnelle}, \citenamefont {Wirth},
  \citenamefont {Chen}, \citenamefont {Goennenwein},\ and\ \citenamefont
  {Felser}}]{Liu2018_NatPhys}%
  \BibitemOpen
  \bibfield  {author} {\bibinfo {author} {\bibfnamefont {E.}~\bibnamefont
  {Liu}}, \bibinfo {author} {\bibfnamefont {Y.}~\bibnamefont {Sun}}, \bibinfo
  {author} {\bibfnamefont {N.}~\bibnamefont {Kumar}}, \bibinfo {author}
  {\bibfnamefont {L.}~\bibnamefont {Muechler}}, \bibinfo {author}
  {\bibfnamefont {A.}~\bibnamefont {Sun}}, \bibinfo {author} {\bibfnamefont
  {L.}~\bibnamefont {Jiao}}, \bibinfo {author} {\bibfnamefont {S.-Y.}\
  \bibnamefont {Yang}}, \bibinfo {author} {\bibfnamefont {D.}~\bibnamefont
  {Liu}}, \bibinfo {author} {\bibfnamefont {A.}~\bibnamefont {Liang}}, \bibinfo
  {author} {\bibfnamefont {Q.}~\bibnamefont {Xu}}, \bibinfo {author}
  {\bibfnamefont {J.}~\bibnamefont {Kroder}}, \bibinfo {author} {\bibfnamefont
  {V.}~\bibnamefont {S{\"u}{\ss}}}, \bibinfo {author} {\bibfnamefont
  {H.}~\bibnamefont {Borrmann}}, \bibinfo {author} {\bibfnamefont
  {C.}~\bibnamefont {Shekhar}}, \bibinfo {author} {\bibfnamefont
  {Z.}~\bibnamefont {Wang}}, \bibinfo {author} {\bibfnamefont {C.}~\bibnamefont
  {Xi}}, \bibinfo {author} {\bibfnamefont {W.}~\bibnamefont {Wang}}, \bibinfo
  {author} {\bibfnamefont {W.}~\bibnamefont {Schnelle}}, \bibinfo {author}
  {\bibfnamefont {S.}~\bibnamefont {Wirth}}, \bibinfo {author} {\bibfnamefont
  {Y.}~\bibnamefont {Chen}}, \bibinfo {author} {\bibfnamefont {S.~T.~B.}\
  \bibnamefont {Goennenwein}}, \ and\ \bibinfo {author} {\bibfnamefont
  {C.}~\bibnamefont {Felser}},\ }\href {\doibase 10.1038/s41567-018-0234-5}
  {\bibfield  {journal} {\bibinfo  {journal} {Nat. Phys.}\ }\textbf {\bibinfo
  {volume} {14}},\ \bibinfo {pages} {1125} (\bibinfo {year}
  {2018})}\BibitemShut {NoStop}%
\bibitem [{\citenamefont {Guin}\ \emph
  {et~al.}(2019{\natexlab{a}})\citenamefont {Guin}, \citenamefont {Manna},
  \citenamefont {Noky}, \citenamefont {Watzman}, \citenamefont {Fu},
  \citenamefont {Kumar}, \citenamefont {Schnelle}, \citenamefont {Shekhar},
  \citenamefont {Sun}, \citenamefont {Gooth},\ and\ \citenamefont
  {Felser}}]{Guin2019}%
  \BibitemOpen
  \bibfield  {author} {\bibinfo {author} {\bibfnamefont {S.~N.}\ \bibnamefont
  {Guin}}, \bibinfo {author} {\bibfnamefont {K.}~\bibnamefont {Manna}},
  \bibinfo {author} {\bibfnamefont {J.}~\bibnamefont {Noky}}, \bibinfo {author}
  {\bibfnamefont {S.~J.}\ \bibnamefont {Watzman}}, \bibinfo {author}
  {\bibfnamefont {C.}~\bibnamefont {Fu}}, \bibinfo {author} {\bibfnamefont
  {N.}~\bibnamefont {Kumar}}, \bibinfo {author} {\bibfnamefont
  {W.}~\bibnamefont {Schnelle}}, \bibinfo {author} {\bibfnamefont
  {C.}~\bibnamefont {Shekhar}}, \bibinfo {author} {\bibfnamefont
  {Y.}~\bibnamefont {Sun}}, \bibinfo {author} {\bibfnamefont {J.}~\bibnamefont
  {Gooth}}, \ and\ \bibinfo {author} {\bibfnamefont {C.}~\bibnamefont
  {Felser}},\ }\href {\doibase 10.1038/s41427-019-0116-z} {\bibfield  {journal}
  {\bibinfo  {journal} {NPG Asia Mater.}\ }\textbf {\bibinfo {volume} {11}},\
  \bibinfo {pages} {16} (\bibinfo {year} {2019}{\natexlab{a}})}\BibitemShut
  {NoStop}%
\bibitem [{\citenamefont {Guin}\ \emph
  {et~al.}(2019{\natexlab{b}})\citenamefont {Guin}, \citenamefont {Vir},
  \citenamefont {Zhang}, \citenamefont {Kumar}, \citenamefont {Watzman},
  \citenamefont {Fu}, \citenamefont {Liu}, \citenamefont {Manna}, \citenamefont
  {Schnelle}, \citenamefont {Gooth}, \citenamefont {Shekhar}, \citenamefont
  {Sun},\ and\ \citenamefont {Felser}}]{Guin_AdvMat_2019}%
  \BibitemOpen
  \bibfield  {author} {\bibinfo {author} {\bibfnamefont {S.~N.}\ \bibnamefont
  {Guin}}, \bibinfo {author} {\bibfnamefont {P.}~\bibnamefont {Vir}}, \bibinfo
  {author} {\bibfnamefont {Y.}~\bibnamefont {Zhang}}, \bibinfo {author}
  {\bibfnamefont {N.}~\bibnamefont {Kumar}}, \bibinfo {author} {\bibfnamefont
  {S.~J.}\ \bibnamefont {Watzman}}, \bibinfo {author} {\bibfnamefont
  {C.}~\bibnamefont {Fu}}, \bibinfo {author} {\bibfnamefont {E.}~\bibnamefont
  {Liu}}, \bibinfo {author} {\bibfnamefont {K.}~\bibnamefont {Manna}}, \bibinfo
  {author} {\bibfnamefont {W.}~\bibnamefont {Schnelle}}, \bibinfo {author}
  {\bibfnamefont {J.}~\bibnamefont {Gooth}}, \bibinfo {author} {\bibfnamefont
  {C.}~\bibnamefont {Shekhar}}, \bibinfo {author} {\bibfnamefont
  {Y.}~\bibnamefont {Sun}}, \ and\ \bibinfo {author} {\bibfnamefont
  {C.}~\bibnamefont {Felser}},\ }\href {\doibase 10.1002/adma.201806622}
  {\bibfield  {journal} {\bibinfo  {journal} {Adv. Mater.}\ }\textbf {\bibinfo
  {volume} {31}},\ \bibinfo {pages} {1806622} (\bibinfo {year}
  {2019}{\natexlab{b}})}\BibitemShut {NoStop}%
\bibitem [{\citenamefont {Xu}\ \emph {et~al.}(2019)\citenamefont {Xu},
  \citenamefont {Phelan},\ and\ \citenamefont
  {Chien}}]{doi:10.1021/acs.nanolett.9b03739}%
  \BibitemOpen
  \bibfield  {author} {\bibinfo {author} {\bibfnamefont {J.}~\bibnamefont
  {Xu}}, \bibinfo {author} {\bibfnamefont {W.~A.}\ \bibnamefont {Phelan}}, \
  and\ \bibinfo {author} {\bibfnamefont {C.-L.}\ \bibnamefont {Chien}},\ }\href
  {\doibase 10.1021/acs.nanolett.9b03739} {\bibfield  {journal} {\bibinfo
  {journal} {Nano Lett.}\ }\textbf {\bibinfo {volume} {19}},\ \bibinfo {pages}
  {8250} (\bibinfo {year} {2019})}\BibitemShut {NoStop}%
\bibitem [{\citenamefont {Sakai}\ \emph {et~al.}(2020)\citenamefont {Sakai},
  \citenamefont {Minami}, \citenamefont {Koretsune}, \citenamefont {Chen},
  \citenamefont {Higo}, \citenamefont {Wang}, \citenamefont {Nomoto},
  \citenamefont {Hirayama}, \citenamefont {Miwa}, \citenamefont
  {Nishio-Hamane}, \citenamefont {Ishii}, \citenamefont {Arita},\ and\
  \citenamefont {Nakatsuji}}]{Fe3X_Nature}%
  \BibitemOpen
  \bibfield  {author} {\bibinfo {author} {\bibfnamefont {A.}~\bibnamefont
  {Sakai}}, \bibinfo {author} {\bibfnamefont {S.}~\bibnamefont {Minami}},
  \bibinfo {author} {\bibfnamefont {T.}~\bibnamefont {Koretsune}}, \bibinfo
  {author} {\bibfnamefont {T.}~\bibnamefont {Chen}}, \bibinfo {author}
  {\bibfnamefont {T.}~\bibnamefont {Higo}}, \bibinfo {author} {\bibfnamefont
  {Y.}~\bibnamefont {Wang}}, \bibinfo {author} {\bibfnamefont {T.}~\bibnamefont
  {Nomoto}}, \bibinfo {author} {\bibfnamefont {M.}~\bibnamefont {Hirayama}},
  \bibinfo {author} {\bibfnamefont {S.}~\bibnamefont {Miwa}}, \bibinfo {author}
  {\bibfnamefont {D.}~\bibnamefont {Nishio-Hamane}}, \bibinfo {author}
  {\bibfnamefont {F.}~\bibnamefont {Ishii}}, \bibinfo {author} {\bibfnamefont
  {R.}~\bibnamefont {Arita}}, \ and\ \bibinfo {author} {\bibfnamefont
  {S.}~\bibnamefont {Nakatsuji}},\ }\href {\doibase 10.1038/s41586-020-2230-z}
  {\bibfield  {journal} {\bibinfo  {journal} {Nature}\ }\textbf {\bibinfo
  {volume} {581}},\ \bibinfo {pages} {53} (\bibinfo {year} {2020})}\BibitemShut
  {NoStop}%
\bibitem [{\citenamefont {Minami}\ \emph {et~al.}(2020)\citenamefont {Minami},
  \citenamefont {Ishii}, \citenamefont {Hirayama}, \citenamefont {Nomoto},
  \citenamefont {Koretsune},\ and\ \citenamefont {Arita}}]{Minami_PRB_2020}%
  \BibitemOpen
  \bibfield  {author} {\bibinfo {author} {\bibfnamefont {S.}~\bibnamefont
  {Minami}}, \bibinfo {author} {\bibfnamefont {F.}~\bibnamefont {Ishii}},
  \bibinfo {author} {\bibfnamefont {M.}~\bibnamefont {Hirayama}}, \bibinfo
  {author} {\bibfnamefont {T.}~\bibnamefont {Nomoto}}, \bibinfo {author}
  {\bibfnamefont {T.}~\bibnamefont {Koretsune}}, \ and\ \bibinfo {author}
  {\bibfnamefont {R.}~\bibnamefont {Arita}},\ }\href {\doibase
  10.1103/PhysRevB.102.205128} {\bibfield  {journal} {\bibinfo  {journal}
  {Phys. Rev. B}\ }\textbf {\bibinfo {volume} {102}},\ \bibinfo {pages}
  {205128} (\bibinfo {year} {2020})}\BibitemShut {NoStop}%
\bibitem [{\citenamefont {Syariati}\ \emph {et~al.}(2020)\citenamefont
  {Syariati}, \citenamefont {Minami}, \citenamefont {Sawahata},\ and\
  \citenamefont {Ishii}}]{Rifky}%
  \BibitemOpen
  \bibfield  {author} {\bibinfo {author} {\bibfnamefont {R.}~\bibnamefont
  {Syariati}}, \bibinfo {author} {\bibfnamefont {S.}~\bibnamefont {Minami}},
  \bibinfo {author} {\bibfnamefont {H.}~\bibnamefont {Sawahata}}, \ and\
  \bibinfo {author} {\bibfnamefont {F.}~\bibnamefont {Ishii}},\ }\href
  {\doibase 10.1063/1.5143474} {\bibfield  {journal} {\bibinfo  {journal} {APL
  Materials}\ }\textbf {\bibinfo {volume} {8}},\ \bibinfo {pages} {041105}
  (\bibinfo {year} {2020})}\BibitemShut {NoStop}%
\bibitem [{\citenamefont {Sumida}\ \emph {et~al.}(2020)\citenamefont {Sumida},
  \citenamefont {Sakuraba}, \citenamefont {Masuda}, \citenamefont {Kono},
  \citenamefont {Kakoki}, \citenamefont {Goto}, \citenamefont {Zhou},
  \citenamefont {Miyamoto}, \citenamefont {Miura}, \citenamefont {Okuda},\ and\
  \citenamefont {Kimura}}]{Sumida_CommunMat_2020}%
  \BibitemOpen
  \bibfield  {author} {\bibinfo {author} {\bibfnamefont {K.}~\bibnamefont
  {Sumida}}, \bibinfo {author} {\bibfnamefont {Y.}~\bibnamefont {Sakuraba}},
  \bibinfo {author} {\bibfnamefont {K.}~\bibnamefont {Masuda}}, \bibinfo
  {author} {\bibfnamefont {T.}~\bibnamefont {Kono}}, \bibinfo {author}
  {\bibfnamefont {M.}~\bibnamefont {Kakoki}}, \bibinfo {author} {\bibfnamefont
  {K.}~\bibnamefont {Goto}}, \bibinfo {author} {\bibfnamefont {W.}~\bibnamefont
  {Zhou}}, \bibinfo {author} {\bibfnamefont {K.}~\bibnamefont {Miyamoto}},
  \bibinfo {author} {\bibfnamefont {Y.}~\bibnamefont {Miura}}, \bibinfo
  {author} {\bibfnamefont {T.}~\bibnamefont {Okuda}}, \ and\ \bibinfo {author}
  {\bibfnamefont {A.}~\bibnamefont {Kimura}},\ }\href {\doibase
  10.1038/s43246-020-00088-w} {\bibfield  {journal} {\bibinfo  {journal}
  {Commun. Mater}\ }\textbf {\bibinfo {volume} {1}},\ \bibinfo {pages} {89}
  (\bibinfo {year} {2020})}\BibitemShut {NoStop}%
\bibitem [{\citenamefont {Yang}\ \emph {et~al.}(2020)\citenamefont {Yang},
  \citenamefont {You}, \citenamefont {Wang}, \citenamefont {Huang},
  \citenamefont {Xi}, \citenamefont {Xu}, \citenamefont {Cao}, \citenamefont
  {Tian}, \citenamefont {Xu}, \citenamefont {Dai},\ and\ \citenamefont
  {Li}}]{Yang_PhysRevMat_2020}%
  \BibitemOpen
  \bibfield  {author} {\bibinfo {author} {\bibfnamefont {H.}~\bibnamefont
  {Yang}}, \bibinfo {author} {\bibfnamefont {W.}~\bibnamefont {You}}, \bibinfo
  {author} {\bibfnamefont {J.}~\bibnamefont {Wang}}, \bibinfo {author}
  {\bibfnamefont {J.}~\bibnamefont {Huang}}, \bibinfo {author} {\bibfnamefont
  {C.}~\bibnamefont {Xi}}, \bibinfo {author} {\bibfnamefont {X.}~\bibnamefont
  {Xu}}, \bibinfo {author} {\bibfnamefont {C.}~\bibnamefont {Cao}}, \bibinfo
  {author} {\bibfnamefont {M.}~\bibnamefont {Tian}}, \bibinfo {author}
  {\bibfnamefont {Z.-A.}\ \bibnamefont {Xu}}, \bibinfo {author} {\bibfnamefont
  {J.}~\bibnamefont {Dai}}, \ and\ \bibinfo {author} {\bibfnamefont
  {Y.}~\bibnamefont {Li}},\ }\href {\doibase 10.1103/PhysRevMaterials.4.024202}
  {\bibfield  {journal} {\bibinfo  {journal} {Phys. Rev. Materials}\ }\textbf
  {\bibinfo {volume} {4}},\ \bibinfo {pages} {024202} (\bibinfo {year}
  {2020})}\BibitemShut {NoStop}%
\bibitem [{\citenamefont {Shi}\ \emph {et~al.}(2020)\citenamefont {Shi},
  \citenamefont {Xu}, \citenamefont {Ma}, \citenamefont {Zhou},\ and\
  \citenamefont {Guo}}]{PhysRevApplied.13.054044}%
  \BibitemOpen
  \bibfield  {author} {\bibinfo {author} {\bibfnamefont {Z.}~\bibnamefont
  {Shi}}, \bibinfo {author} {\bibfnamefont {S.-J.}\ \bibnamefont {Xu}},
  \bibinfo {author} {\bibfnamefont {L.}~\bibnamefont {Ma}}, \bibinfo {author}
  {\bibfnamefont {S.-M.}\ \bibnamefont {Zhou}}, \ and\ \bibinfo {author}
  {\bibfnamefont {G.-Y.}\ \bibnamefont {Guo}},\ }\href {\doibase
  10.1103/PhysRevApplied.13.054044} {\bibfield  {journal} {\bibinfo  {journal}
  {Phys. Rev. Applied}\ }\textbf {\bibinfo {volume} {13}},\ \bibinfo {pages}
  {054044} (\bibinfo {year} {2020})}\BibitemShut {NoStop}%
\bibitem [{\citenamefont {Hirschberger}\ \emph {et~al.}(2020)\citenamefont
  {Hirschberger}, \citenamefont {Spitz}, \citenamefont {Nomoto}, \citenamefont
  {Kurumaji}, \citenamefont {Gao}, \citenamefont {Masell}, \citenamefont
  {Nakajima}, \citenamefont {Kikkawa}, \citenamefont {Yamasaki}, \citenamefont
  {Sagayama}, \citenamefont {Nakao}, \citenamefont {Taguchi}, \citenamefont
  {Arita}, \citenamefont {Arima},\ and\ \citenamefont {Tokura}}]{Max_PRL_2020}%
  \BibitemOpen
  \bibfield  {author} {\bibinfo {author} {\bibfnamefont {M.}~\bibnamefont
  {Hirschberger}}, \bibinfo {author} {\bibfnamefont {L.}~\bibnamefont {Spitz}},
  \bibinfo {author} {\bibfnamefont {T.}~\bibnamefont {Nomoto}}, \bibinfo
  {author} {\bibfnamefont {T.}~\bibnamefont {Kurumaji}}, \bibinfo {author}
  {\bibfnamefont {S.}~\bibnamefont {Gao}}, \bibinfo {author} {\bibfnamefont
  {J.}~\bibnamefont {Masell}}, \bibinfo {author} {\bibfnamefont
  {T.}~\bibnamefont {Nakajima}}, \bibinfo {author} {\bibfnamefont
  {A.}~\bibnamefont {Kikkawa}}, \bibinfo {author} {\bibfnamefont
  {Y.}~\bibnamefont {Yamasaki}}, \bibinfo {author} {\bibfnamefont
  {H.}~\bibnamefont {Sagayama}}, \bibinfo {author} {\bibfnamefont
  {H.}~\bibnamefont {Nakao}}, \bibinfo {author} {\bibfnamefont
  {Y.}~\bibnamefont {Taguchi}}, \bibinfo {author} {\bibfnamefont
  {R.}~\bibnamefont {Arita}}, \bibinfo {author} {\bibfnamefont {T.-h.}\
  \bibnamefont {Arima}}, \ and\ \bibinfo {author} {\bibfnamefont
  {Y.}~\bibnamefont {Tokura}},\ }\href {\doibase
  10.1103/PhysRevLett.125.076602} {\bibfield  {journal} {\bibinfo  {journal}
  {Phys. Rev. Lett.}\ }\textbf {\bibinfo {volume} {125}},\ \bibinfo {pages}
  {076602} (\bibinfo {year} {2020})}\BibitemShut {NoStop}%
\bibitem [{\citenamefont {Chen}\ \emph {et~al.}(2021)\citenamefont {Chen},
  \citenamefont {Tomita}, \citenamefont {Minami}, \citenamefont {Fu},
  \citenamefont {Koretsune}, \citenamefont {Kitatani}, \citenamefont
  {Muhammad}, \citenamefont {Nishio-Hamane}, \citenamefont {Ishii},
  \citenamefont {Ishii}, \citenamefont {Arita},\ and\ \citenamefont
  {Nakatsuji}}]{Taishi_NatCom_2021}%
  \BibitemOpen
  \bibfield  {author} {\bibinfo {author} {\bibfnamefont {T.}~\bibnamefont
  {Chen}}, \bibinfo {author} {\bibfnamefont {T.}~\bibnamefont {Tomita}},
  \bibinfo {author} {\bibfnamefont {S.}~\bibnamefont {Minami}}, \bibinfo
  {author} {\bibfnamefont {M.}~\bibnamefont {Fu}}, \bibinfo {author}
  {\bibfnamefont {T.}~\bibnamefont {Koretsune}}, \bibinfo {author}
  {\bibfnamefont {M.}~\bibnamefont {Kitatani}}, \bibinfo {author}
  {\bibfnamefont {I.}~\bibnamefont {Muhammad}}, \bibinfo {author}
  {\bibfnamefont {D.}~\bibnamefont {Nishio-Hamane}}, \bibinfo {author}
  {\bibfnamefont {R.}~\bibnamefont {Ishii}}, \bibinfo {author} {\bibfnamefont
  {F.}~\bibnamefont {Ishii}}, \bibinfo {author} {\bibfnamefont
  {R.}~\bibnamefont {Arita}}, \ and\ \bibinfo {author} {\bibfnamefont
  {S.}~\bibnamefont {Nakatsuji}},\ }\href {\doibase 10.1038/s41467-020-20838-1}
  {\bibfield  {journal} {\bibinfo  {journal} {Nat. Commun.}\ }\textbf {\bibinfo
  {volume} {12}},\ \bibinfo {pages} {572} (\bibinfo {year} {2021})}\BibitemShut
  {NoStop}%
\bibitem [{\citenamefont {Asaba}\ \emph {et~al.}(2021)\citenamefont {Asaba},
  \citenamefont {Ivanov}, \citenamefont {Thomas}, \citenamefont {Savrasov},
  \citenamefont {Thompson}, \citenamefont {Bauer},\ and\ \citenamefont
  {Ronning}}]{TAsaba_SicAdv_2021}%
  \BibitemOpen
  \bibfield  {author} {\bibinfo {author} {\bibfnamefont {T.}~\bibnamefont
  {Asaba}}, \bibinfo {author} {\bibfnamefont {V.}~\bibnamefont {Ivanov}},
  \bibinfo {author} {\bibfnamefont {S.~M.}\ \bibnamefont {Thomas}}, \bibinfo
  {author} {\bibfnamefont {S.~Y.}\ \bibnamefont {Savrasov}}, \bibinfo {author}
  {\bibfnamefont {J.~D.}\ \bibnamefont {Thompson}}, \bibinfo {author}
  {\bibfnamefont {E.~D.}\ \bibnamefont {Bauer}}, \ and\ \bibinfo {author}
  {\bibfnamefont {F.}~\bibnamefont {Ronning}},\ }\href {\doibase
  10.1126/sciadv.abf1467} {\bibfield  {journal} {\bibinfo  {journal} {Sci.
  Adv.}\ }\textbf {\bibinfo {volume} {7}},\ \bibinfo {pages} {eabf1467}
  (\bibinfo {year} {2021})}\BibitemShut {NoStop}%
\bibitem [{\citenamefont {He}\ \emph {et~al.}(2021)\citenamefont {He},
  \citenamefont {Şahin}, \citenamefont {Boona}, \citenamefont {Sales},
  \citenamefont {Pan}, \citenamefont {Felser}, \citenamefont {Flatté},\ and\
  \citenamefont {Heremans}}]{BinHe_Joule_2021}%
  \BibitemOpen
  \bibfield  {author} {\bibinfo {author} {\bibfnamefont {B.}~\bibnamefont
  {He}}, \bibinfo {author} {\bibfnamefont {C.}~\bibnamefont {Şahin}}, \bibinfo
  {author} {\bibfnamefont {S.~R.}\ \bibnamefont {Boona}}, \bibinfo {author}
  {\bibfnamefont {B.~C.}\ \bibnamefont {Sales}}, \bibinfo {author}
  {\bibfnamefont {Y.}~\bibnamefont {Pan}}, \bibinfo {author} {\bibfnamefont
  {C.}~\bibnamefont {Felser}}, \bibinfo {author} {\bibfnamefont {M.~E.}\
  \bibnamefont {Flatté}}, \ and\ \bibinfo {author} {\bibfnamefont {J.~P.}\
  \bibnamefont {Heremans}},\ }\href {\doibase
  https://doi.org/10.1016/j.joule.2021.08.007} {\bibfield  {journal} {\bibinfo
  {journal} {Joule}\ }\textbf {\bibinfo {volume} {5}},\ \bibinfo {pages} {3057}
  (\bibinfo {year} {2021})}\BibitemShut {NoStop}%
\bibitem [{\citenamefont {Nakamura}\ \emph {et~al.}(2021)\citenamefont
  {Nakamura}, \citenamefont {Minami}, \citenamefont {Tomita}, \citenamefont
  {Nugroho},\ and\ \citenamefont {Nakatsuji}}]{Nakamura_PRBL_2021}%
  \BibitemOpen
  \bibfield  {author} {\bibinfo {author} {\bibfnamefont {H.}~\bibnamefont
  {Nakamura}}, \bibinfo {author} {\bibfnamefont {S.}~\bibnamefont {Minami}},
  \bibinfo {author} {\bibfnamefont {T.}~\bibnamefont {Tomita}}, \bibinfo
  {author} {\bibfnamefont {A.~A.}\ \bibnamefont {Nugroho}}, \ and\ \bibinfo
  {author} {\bibfnamefont {S.}~\bibnamefont {Nakatsuji}},\ }\href {\doibase
  10.1103/PhysRevB.104.L161114} {\bibfield  {journal} {\bibinfo  {journal}
  {Phys. Rev. B}\ }\textbf {\bibinfo {volume} {104}},\ \bibinfo {pages}
  {L161114} (\bibinfo {year} {2021})}\BibitemShut {NoStop}%
\bibitem [{\citenamefont {Chen}\ \emph {et~al.}(2022)\citenamefont {Chen},
  \citenamefont {Minami}, \citenamefont {Sakai}, \citenamefont {Wang},
  \citenamefont {Feng}, \citenamefont {Nomoto}, \citenamefont {Hirayama},
  \citenamefont {Ishii}, \citenamefont {Koretsune}, \citenamefont {Arita},\
  and\ \citenamefont {Nakatsuji}}]{Taishi_SicAdv_2022}%
  \BibitemOpen
  \bibfield  {author} {\bibinfo {author} {\bibfnamefont {T.}~\bibnamefont
  {Chen}}, \bibinfo {author} {\bibfnamefont {S.}~\bibnamefont {Minami}},
  \bibinfo {author} {\bibfnamefont {A.}~\bibnamefont {Sakai}}, \bibinfo
  {author} {\bibfnamefont {Y.}~\bibnamefont {Wang}}, \bibinfo {author}
  {\bibfnamefont {Z.}~\bibnamefont {Feng}}, \bibinfo {author} {\bibfnamefont
  {T.}~\bibnamefont {Nomoto}}, \bibinfo {author} {\bibfnamefont
  {M.}~\bibnamefont {Hirayama}}, \bibinfo {author} {\bibfnamefont
  {R.}~\bibnamefont {Ishii}}, \bibinfo {author} {\bibfnamefont
  {T.}~\bibnamefont {Koretsune}}, \bibinfo {author} {\bibfnamefont
  {R.}~\bibnamefont {Arita}}, \ and\ \bibinfo {author} {\bibfnamefont
  {S.}~\bibnamefont {Nakatsuji}},\ }\href {\doibase 10.1126/sciadv.abk1480}
  {\bibfield  {journal} {\bibinfo  {journal} {Sci. Adv.}\ }\textbf {\bibinfo
  {volume} {8}},\ \bibinfo {pages} {eabk1480} (\bibinfo {year}
  {2022})}\BibitemShut {NoStop}%
\bibitem [{\citenamefont {Mizuta}\ and\ \citenamefont
  {Ishii}(2014)}]{YMizuta_JPS}%
  \BibitemOpen
  \bibfield  {author} {\bibinfo {author} {\bibfnamefont {Y.~P.}\ \bibnamefont
  {Mizuta}}\ and\ \bibinfo {author} {\bibfnamefont {F.}~\bibnamefont {Ishii}},\
  }\href {\doibase 10.7566/JPSCP.3.017035} {\bibfield  {journal} {\bibinfo
  {journal} {JPS Conf. Proc.}\ }\textbf {\bibinfo {volume} {3}},\ \bibinfo
  {pages} {017035} (\bibinfo {year} {2014})}\BibitemShut {NoStop}%
\bibitem [{\citenamefont {Wang}\ \emph {et~al.}(2006)\citenamefont {Wang},
  \citenamefont {Yates}, \citenamefont {Souza},\ and\ \citenamefont
  {Vanderbilt}}]{Vanderbilt}%
  \BibitemOpen
  \bibfield  {author} {\bibinfo {author} {\bibfnamefont {X.}~\bibnamefont
  {Wang}}, \bibinfo {author} {\bibfnamefont {J.~R.}\ \bibnamefont {Yates}},
  \bibinfo {author} {\bibfnamefont {I.}~\bibnamefont {Souza}}, \ and\ \bibinfo
  {author} {\bibfnamefont {D.}~\bibnamefont {Vanderbilt}},\ }\href {\doibase
  10.1103/PhysRevB.74.195118} {\bibfield  {journal} {\bibinfo  {journal} {Phys.
  Rev. B}\ }\textbf {\bibinfo {volume} {74}},\ \bibinfo {pages} {195118}
  (\bibinfo {year} {2006})}\BibitemShut {NoStop}%
\bibitem [{\citenamefont {Lopez}\ \emph {et~al.}(2012)\citenamefont {Lopez},
  \citenamefont {Vanderbilt}, \citenamefont {Thonhauser},\ and\ \citenamefont
  {Souza}}]{Lopez}%
  \BibitemOpen
  \bibfield  {author} {\bibinfo {author} {\bibfnamefont {M.~G.}\ \bibnamefont
  {Lopez}}, \bibinfo {author} {\bibfnamefont {D.}~\bibnamefont {Vanderbilt}},
  \bibinfo {author} {\bibfnamefont {T.}~\bibnamefont {Thonhauser}}, \ and\
  \bibinfo {author} {\bibfnamefont {I.}~\bibnamefont {Souza}},\ }\href
  {\doibase 10.1103/PhysRevB.85.014435} {\bibfield  {journal} {\bibinfo
  {journal} {Phys. Rev. B}\ }\textbf {\bibinfo {volume} {85}},\ \bibinfo
  {pages} {014435} (\bibinfo {year} {2012})}\BibitemShut {NoStop}%
\bibitem [{\citenamefont {Tsirkin}(2021)}]{Tsirkin}%
  \BibitemOpen
  \bibfield  {author} {\bibinfo {author} {\bibfnamefont {S.~S.}\ \bibnamefont
  {Tsirkin}},\ }\href {https://doi.org/10.1038/s41524-021-00498-5} {\bibfield
  {journal} {\bibinfo  {journal} {npj Computational Materials}\ }\textbf
  {\bibinfo {volume} {7}},\ \bibinfo {pages} {33} (\bibinfo {year}
  {2021})}\BibitemShut {NoStop}%
\bibitem [{\citenamefont {Yao}\ \emph {et~al.}(2004)\citenamefont {Yao},
  \citenamefont {Kleinman}, \citenamefont {MacDonald}, \citenamefont {Sinova},
  \citenamefont {Jungwirth}, \citenamefont {Wang}, \citenamefont {Wang},\ and\
  \citenamefont {Niu}}]{YYao}%
  \BibitemOpen
  \bibfield  {author} {\bibinfo {author} {\bibfnamefont {Y.}~\bibnamefont
  {Yao}}, \bibinfo {author} {\bibfnamefont {L.}~\bibnamefont {Kleinman}},
  \bibinfo {author} {\bibfnamefont {A.~H.}\ \bibnamefont {MacDonald}}, \bibinfo
  {author} {\bibfnamefont {J.}~\bibnamefont {Sinova}}, \bibinfo {author}
  {\bibfnamefont {T.}~\bibnamefont {Jungwirth}}, \bibinfo {author}
  {\bibfnamefont {D.-s.}\ \bibnamefont {Wang}}, \bibinfo {author}
  {\bibfnamefont {E.}~\bibnamefont {Wang}}, \ and\ \bibinfo {author}
  {\bibfnamefont {Q.}~\bibnamefont {Niu}},\ }\href@noop {} {\bibfield
  {journal} {\bibinfo  {journal} {Phys. Rev. Lett.}\ }\textbf {\bibinfo
  {volume} {92}},\ \bibinfo {pages} {037204} (\bibinfo {year}
  {2004})}\BibitemShut {NoStop}%
\bibitem [{\citenamefont {Pizzi}\ \emph {et~al.}(2020)\citenamefont {Pizzi},
  \citenamefont {Vitale}, \citenamefont {Arita}, \citenamefont {Blügel},
  \citenamefont {Freimuth}, \citenamefont {G{\'{e}}ranton}, \citenamefont
  {Gibertini}, \citenamefont {Gresch}, \citenamefont {Johnson}, \citenamefont
  {Koretsune}, \citenamefont {Iba{\~{n}}ez-Azpiroz}, \citenamefont {Lee},
  \citenamefont {Lihm}, \citenamefont {Marchand}, \citenamefont {Marrazzo},
  \citenamefont {Mokrousov}, \citenamefont {Mustafa}, \citenamefont {Nohara},
  \citenamefont {Nomura}, \citenamefont {Paulatto}, \citenamefont
  {Ponc{\'{e}}}, \citenamefont {Ponweiser}, \citenamefont {Qiao}, \citenamefont
  {Thöle}, \citenamefont {Tsirkin}, \citenamefont {Wierzbowska}, \citenamefont
  {Marzari}, \citenamefont {Vanderbilt}, \citenamefont {Souza}, \citenamefont
  {Mostofi},\ and\ \citenamefont {Yates}}]{Wannier90}%
  \BibitemOpen
  \bibfield  {author} {\bibinfo {author} {\bibfnamefont {G.}~\bibnamefont
  {Pizzi}}, \bibinfo {author} {\bibfnamefont {V.}~\bibnamefont {Vitale}},
  \bibinfo {author} {\bibfnamefont {R.}~\bibnamefont {Arita}}, \bibinfo
  {author} {\bibfnamefont {S.}~\bibnamefont {Blügel}}, \bibinfo {author}
  {\bibfnamefont {F.}~\bibnamefont {Freimuth}}, \bibinfo {author}
  {\bibfnamefont {G.}~\bibnamefont {G{\'{e}}ranton}}, \bibinfo {author}
  {\bibfnamefont {M.}~\bibnamefont {Gibertini}}, \bibinfo {author}
  {\bibfnamefont {D.}~\bibnamefont {Gresch}}, \bibinfo {author} {\bibfnamefont
  {C.}~\bibnamefont {Johnson}}, \bibinfo {author} {\bibfnamefont
  {T.}~\bibnamefont {Koretsune}}, \bibinfo {author} {\bibfnamefont
  {J.}~\bibnamefont {Iba{\~{n}}ez-Azpiroz}}, \bibinfo {author} {\bibfnamefont
  {H.}~\bibnamefont {Lee}}, \bibinfo {author} {\bibfnamefont {J.-M.}\
  \bibnamefont {Lihm}}, \bibinfo {author} {\bibfnamefont {D.}~\bibnamefont
  {Marchand}}, \bibinfo {author} {\bibfnamefont {A.}~\bibnamefont {Marrazzo}},
  \bibinfo {author} {\bibfnamefont {Y.}~\bibnamefont {Mokrousov}}, \bibinfo
  {author} {\bibfnamefont {J.~I.}\ \bibnamefont {Mustafa}}, \bibinfo {author}
  {\bibfnamefont {Y.}~\bibnamefont {Nohara}}, \bibinfo {author} {\bibfnamefont
  {Y.}~\bibnamefont {Nomura}}, \bibinfo {author} {\bibfnamefont
  {L.}~\bibnamefont {Paulatto}}, \bibinfo {author} {\bibfnamefont
  {S.}~\bibnamefont {Ponc{\'{e}}}}, \bibinfo {author} {\bibfnamefont
  {T.}~\bibnamefont {Ponweiser}}, \bibinfo {author} {\bibfnamefont
  {J.}~\bibnamefont {Qiao}}, \bibinfo {author} {\bibfnamefont {F.}~\bibnamefont
  {Thöle}}, \bibinfo {author} {\bibfnamefont {S.~S.}\ \bibnamefont {Tsirkin}},
  \bibinfo {author} {\bibfnamefont {M.}~\bibnamefont {Wierzbowska}}, \bibinfo
  {author} {\bibfnamefont {N.}~\bibnamefont {Marzari}}, \bibinfo {author}
  {\bibfnamefont {D.}~\bibnamefont {Vanderbilt}}, \bibinfo {author}
  {\bibfnamefont {I.}~\bibnamefont {Souza}}, \bibinfo {author} {\bibfnamefont
  {A.~A.}\ \bibnamefont {Mostofi}}, \ and\ \bibinfo {author} {\bibfnamefont
  {J.~R.}\ \bibnamefont {Yates}},\ }\href {\doibase 10.1088/1361-648x/ab51ff}
  {\bibfield  {journal} {\bibinfo  {journal} {Journal of Physics: Condensed
  Matter}\ }\textbf {\bibinfo {volume} {32}},\ \bibinfo {pages} {165902}
  (\bibinfo {year} {2020})}\BibitemShut {NoStop}%
\bibitem [{\citenamefont {Fukui}\ \emph {et~al.}(2005)\citenamefont {Fukui},
  \citenamefont {Hatsugai},\ and\ \citenamefont {Suzuki}}]{FHS}%
  \BibitemOpen
  \bibfield  {author} {\bibinfo {author} {\bibfnamefont {T.}~\bibnamefont
  {Fukui}}, \bibinfo {author} {\bibfnamefont {Y.}~\bibnamefont {Hatsugai}}, \
  and\ \bibinfo {author} {\bibfnamefont {H.}~\bibnamefont {Suzuki}},\ }\href
  {\doibase 10.1143/JPSJ.74.1674} {\bibfield  {journal} {\bibinfo  {journal}
  {Journal of the Physical Society of Japan}\ }\textbf {\bibinfo {volume}
  {74}},\ \bibinfo {pages} {1674} (\bibinfo {year} {2005})}\BibitemShut
  {NoStop}%
\bibitem [{\citenamefont {Fukui}\ and\ \citenamefont {Hatsugai}(2007)}]{FH}%
  \BibitemOpen
  \bibfield  {author} {\bibinfo {author} {\bibfnamefont {T.}~\bibnamefont
  {Fukui}}\ and\ \bibinfo {author} {\bibfnamefont {Y.}~\bibnamefont
  {Hatsugai}},\ }\href {\doibase 10.1143/JPSJ.76.053702} {\bibfield  {journal}
  {\bibinfo  {journal} {Journal of the Physical Society of Japan}\ }\textbf
  {\bibinfo {volume} {76}},\ \bibinfo {pages} {053702} (\bibinfo {year}
  {2007})}\BibitemShut {NoStop}%
\bibitem [{\citenamefont {Feng}\ \emph {et~al.}(2012)\citenamefont {Feng},
  \citenamefont {Wen}, \citenamefont {Zhou}, \citenamefont {Xiao},\ and\
  \citenamefont {Yao}}]{FENG20121849}%
  \BibitemOpen
  \bibfield  {author} {\bibinfo {author} {\bibfnamefont {W.}~\bibnamefont
  {Feng}}, \bibinfo {author} {\bibfnamefont {J.}~\bibnamefont {Wen}}, \bibinfo
  {author} {\bibfnamefont {J.}~\bibnamefont {Zhou}}, \bibinfo {author}
  {\bibfnamefont {D.}~\bibnamefont {Xiao}}, \ and\ \bibinfo {author}
  {\bibfnamefont {Y.}~\bibnamefont {Yao}},\ }\href {\doibase
  https://doi.org/10.1016/j.cpc.2012.04.001} {\bibfield  {journal} {\bibinfo
  {journal} {Computer Physics Communications}\ }\textbf {\bibinfo {volume}
  {183}},\ \bibinfo {pages} {1849} (\bibinfo {year} {2012})}\BibitemShut
  {NoStop}%
\bibitem [{\citenamefont {{T. Ozaki et al.}}()}]{openmx}%
  \BibitemOpen
  \bibfield  {author} {\bibinfo {author} {\bibnamefont {{T. Ozaki et al.}}},\
  }\href@noop {} {\enquote {\bibinfo {title} {{OpenMX: Open source package for
  Material eXplorer}},}\ }\bibinfo {howpublished}
  {{\url{http://www.openmx-square.org/.}}}\BibitemShut {Stop}%
\bibitem [{\citenamefont {Lee}\ \emph {et~al.}(2018)\citenamefont {Lee},
  \citenamefont {Lee}, \citenamefont {Fukuda},\ and\ \citenamefont
  {Ozaki}}]{CCLee}%
  \BibitemOpen
  \bibfield  {author} {\bibinfo {author} {\bibfnamefont {C.-C.}\ \bibnamefont
  {Lee}}, \bibinfo {author} {\bibfnamefont {Y.-T.}\ \bibnamefont {Lee}},
  \bibinfo {author} {\bibfnamefont {M.}~\bibnamefont {Fukuda}}, \ and\ \bibinfo
  {author} {\bibfnamefont {T.}~\bibnamefont {Ozaki}},\ }\href {\doibase
  10.1103/PhysRevB.98.115115} {\bibfield  {journal} {\bibinfo  {journal} {Phys.
  Rev. B}\ }\textbf {\bibinfo {volume} {98}},\ \bibinfo {pages} {115115}
  (\bibinfo {year} {2018})}\BibitemShut {NoStop}%
\bibitem [{\citenamefont {Fuh}\ and\ \citenamefont {Guo}(2011)}]{Guo_PRB_2011}%
  \BibitemOpen
  \bibfield  {author} {\bibinfo {author} {\bibfnamefont {H.-R.}\ \bibnamefont
  {Fuh}}\ and\ \bibinfo {author} {\bibfnamefont {G.-Y.}\ \bibnamefont {Guo}},\
  }\href {\doibase 10.1103/PhysRevB.84.144427} {\bibfield  {journal} {\bibinfo
  {journal} {Phys. Rev. B}\ }\textbf {\bibinfo {volume} {84}},\ \bibinfo
  {pages} {144427} (\bibinfo {year} {2011})}\BibitemShut {NoStop}%
\bibitem [{\citenamefont {Weischenberg}\ \emph {et~al.}(2013)\citenamefont
  {Weischenberg}, \citenamefont {Freimuth}, \citenamefont {Bl\"ugel},\ and\
  \citenamefont {Mokrousov}}]{Weischenberg_PRB_2013}%
  \BibitemOpen
  \bibfield  {author} {\bibinfo {author} {\bibfnamefont {J.}~\bibnamefont
  {Weischenberg}}, \bibinfo {author} {\bibfnamefont {F.}~\bibnamefont
  {Freimuth}}, \bibinfo {author} {\bibfnamefont {S.}~\bibnamefont {Bl\"ugel}},
  \ and\ \bibinfo {author} {\bibfnamefont {Y.}~\bibnamefont {Mokrousov}},\
  }\href {\doibase 10.1103/PhysRevB.87.060406} {\bibfield  {journal} {\bibinfo
  {journal} {Phys. Rev. B}\ }\textbf {\bibinfo {volume} {87}},\ \bibinfo
  {pages} {060406} (\bibinfo {year} {2013})}\BibitemShut {NoStop}%
\bibitem [{Note1()}]{Note1}%
  \BibitemOpen
  \bibinfo {note} {\protect \url
  {https://github.com/hikaruri/OMXsigmaxy}}\BibitemShut {NoStop}%
\bibitem [{\citenamefont {King-Smith}\ and\ \citenamefont
  {Vanderbilt}(1993)}]{PhysRevB.47.1651}%
  \BibitemOpen
  \bibfield  {author} {\bibinfo {author} {\bibfnamefont {R.~D.}\ \bibnamefont
  {King-Smith}}\ and\ \bibinfo {author} {\bibfnamefont {D.}~\bibnamefont
  {Vanderbilt}},\ }\href {\doibase 10.1103/PhysRevB.47.1651} {\bibfield
  {journal} {\bibinfo  {journal} {Phys. Rev. B}\ }\textbf {\bibinfo {volume}
  {47}},\ \bibinfo {pages} {1651} (\bibinfo {year} {1993})}\BibitemShut
  {NoStop}%
\bibitem [{\citenamefont {Perdew}\ \emph {et~al.}(1996)\citenamefont {Perdew},
  \citenamefont {Burke},\ and\ \citenamefont
  {Ernzerhof}}]{PhysRevLett.77.3865}%
  \BibitemOpen
  \bibfield  {author} {\bibinfo {author} {\bibfnamefont {J.~P.}\ \bibnamefont
  {Perdew}}, \bibinfo {author} {\bibfnamefont {K.}~\bibnamefont {Burke}}, \
  and\ \bibinfo {author} {\bibfnamefont {M.}~\bibnamefont {Ernzerhof}},\ }\href
  {\doibase 10.1103/PhysRevLett.77.3865} {\bibfield  {journal} {\bibinfo
  {journal} {Phys. Rev. Lett.}\ }\textbf {\bibinfo {volume} {77}},\ \bibinfo
  {pages} {3865} (\bibinfo {year} {1996})}\BibitemShut {NoStop}%
\bibitem [{\citenamefont {Hamann}\ \emph {et~al.}(1979)\citenamefont {Hamann},
  \citenamefont {Schl\"uter},\ and\ \citenamefont
  {Chiang}}]{PhysRevLett.43.1494}%
  \BibitemOpen
  \bibfield  {author} {\bibinfo {author} {\bibfnamefont {D.~R.}\ \bibnamefont
  {Hamann}}, \bibinfo {author} {\bibfnamefont {M.}~\bibnamefont {Schl\"uter}},
  \ and\ \bibinfo {author} {\bibfnamefont {C.}~\bibnamefont {Chiang}},\ }\href
  {\doibase 10.1103/PhysRevLett.43.1494} {\bibfield  {journal} {\bibinfo
  {journal} {Phys. Rev. Lett.}\ }\textbf {\bibinfo {volume} {43}},\ \bibinfo
  {pages} {1494} (\bibinfo {year} {1979})}\BibitemShut {NoStop}%
\bibitem [{\citenamefont {Ozaki}(2003)}]{OzakiPRB67}%
  \BibitemOpen
  \bibfield  {author} {\bibinfo {author} {\bibfnamefont {T.}~\bibnamefont
  {Ozaki}},\ }\href {\doibase 10.1103/PhysRevB.67.155108} {\bibfield  {journal}
  {\bibinfo  {journal} {Phys. Rev. B}\ }\textbf {\bibinfo {volume} {67}},\
  \bibinfo {pages} {155108} (\bibinfo {year} {2003})}\BibitemShut {NoStop}%
\bibitem [{\citenamefont {Ozaki}\ and\ \citenamefont
  {Kino}(2004)}]{OzakiKinoPRB69}%
  \BibitemOpen
  \bibfield  {author} {\bibinfo {author} {\bibfnamefont {T.}~\bibnamefont
  {Ozaki}}\ and\ \bibinfo {author} {\bibfnamefont {H.}~\bibnamefont {Kino}},\
  }\href {\doibase 10.1103/PhysRevB.69.195113} {\bibfield  {journal} {\bibinfo
  {journal} {Phys. Rev. B}\ }\textbf {\bibinfo {volume} {69}},\ \bibinfo
  {pages} {195113} (\bibinfo {year} {2004})}\BibitemShut {NoStop}%
\bibitem [{\citenamefont {Theurich}\ and\ \citenamefont
  {Hill}(2001)}]{PhysRevB.64.073106}%
  \BibitemOpen
  \bibfield  {author} {\bibinfo {author} {\bibfnamefont {G.}~\bibnamefont
  {Theurich}}\ and\ \bibinfo {author} {\bibfnamefont {N.~A.}\ \bibnamefont
  {Hill}},\ }\href {\doibase 10.1103/PhysRevB.64.073106} {\bibfield  {journal}
  {\bibinfo  {journal} {Phys. Rev. B}\ }\textbf {\bibinfo {volume} {64}},\
  \bibinfo {pages} {073106} (\bibinfo {year} {2001})}\BibitemShut {NoStop}%
\bibitem [{\citenamefont {Kulish}\ and\ \citenamefont
  {Huang}(2017)}]{C7TC02664A}%
  \BibitemOpen
  \bibfield  {author} {\bibinfo {author} {\bibfnamefont {V.~V.}\ \bibnamefont
  {Kulish}}\ and\ \bibinfo {author} {\bibfnamefont {W.}~\bibnamefont {Huang}},\
  }\href {\doibase 10.1039/C7TC02664A} {\bibfield  {journal} {\bibinfo
  {journal} {J. Mater. Chem. C}\ }\textbf {\bibinfo {volume} {5}},\ \bibinfo
  {pages} {8734} (\bibinfo {year} {2017})}\BibitemShut {NoStop}%
\bibitem [{\citenamefont {Zheng}\ \emph {et~al.}(2017)\citenamefont {Zheng},
  \citenamefont {Zheng}, \citenamefont {Wang}, \citenamefont {Han},\ and\
  \citenamefont {Yan}}]{ZHENG2017184}%
  \BibitemOpen
  \bibfield  {author} {\bibinfo {author} {\bibfnamefont {H.}~\bibnamefont
  {Zheng}}, \bibinfo {author} {\bibfnamefont {J.}~\bibnamefont {Zheng}},
  \bibinfo {author} {\bibfnamefont {C.}~\bibnamefont {Wang}}, \bibinfo {author}
  {\bibfnamefont {H.}~\bibnamefont {Han}}, \ and\ \bibinfo {author}
  {\bibfnamefont {Y.}~\bibnamefont {Yan}},\ }\href {\doibase
  https://doi.org/10.1016/j.jmmm.2017.08.005} {\bibfield  {journal} {\bibinfo
  {journal} {Journal of Magnetism and Magnetic Materials}\ }\textbf {\bibinfo
  {volume} {444}},\ \bibinfo {pages} {184} (\bibinfo {year}
  {2017})}\BibitemShut {NoStop}%
\bibitem [{\citenamefont {K\"ubler}\ and\ \citenamefont
  {Felser}(2012)}]{PhysRevB.85.012405}%
  \BibitemOpen
  \bibfield  {author} {\bibinfo {author} {\bibfnamefont {J.}~\bibnamefont
  {K\"ubler}}\ and\ \bibinfo {author} {\bibfnamefont {C.}~\bibnamefont
  {Felser}},\ }\href {\doibase 10.1103/PhysRevB.85.012405} {\bibfield
  {journal} {\bibinfo  {journal} {Phys. Rev. B}\ }\textbf {\bibinfo {volume}
  {85}},\ \bibinfo {pages} {012405} (\bibinfo {year} {2012})}\BibitemShut
  {NoStop}%
\bibitem [{\citenamefont {Dheer}(1967)}]{Dheer}%
  \BibitemOpen
  \bibfield  {author} {\bibinfo {author} {\bibfnamefont {P.~N.}\ \bibnamefont
  {Dheer}},\ }\href {\doibase 10.1103/PhysRev.156.637} {\bibfield  {journal}
  {\bibinfo  {journal} {Phys. Rev.}\ }\textbf {\bibinfo {volume} {156}},\
  \bibinfo {pages} {637} (\bibinfo {year} {1967})}\BibitemShut {NoStop}%
\bibitem [{\citenamefont {Momma}\ and\ \citenamefont
  {Izumi}(2011)}]{Momma:db5098}%
  \BibitemOpen
  \bibfield  {author} {\bibinfo {author} {\bibfnamefont {K.}~\bibnamefont
  {Momma}}\ and\ \bibinfo {author} {\bibfnamefont {F.}~\bibnamefont {Izumi}},\
  }\href {\doibase 10.1107/S0021889811038970} {\bibfield  {journal} {\bibinfo
  {journal} {Journal of Applied Crystallography}\ }\textbf {\bibinfo {volume}
  {44}},\ \bibinfo {pages} {1272} (\bibinfo {year} {2011})}\BibitemShut
  {NoStop}%
\bibitem [{\citenamefont {Gos\'albez-Mart\'{\i}nez}\ \emph
  {et~al.}(2015)\citenamefont {Gos\'albez-Mart\'{\i}nez}, \citenamefont
  {Souza},\ and\ \citenamefont {Vanderbilt}}]{PhysRevB.92.085138}%
  \BibitemOpen
  \bibfield  {author} {\bibinfo {author} {\bibfnamefont {D.}~\bibnamefont
  {Gos\'albez-Mart\'{\i}nez}}, \bibinfo {author} {\bibfnamefont
  {I.}~\bibnamefont {Souza}}, \ and\ \bibinfo {author} {\bibfnamefont
  {D.}~\bibnamefont {Vanderbilt}},\ }\href {\doibase
  10.1103/PhysRevB.92.085138} {\bibfield  {journal} {\bibinfo  {journal} {Phys.
  Rev. B}\ }\textbf {\bibinfo {volume} {92}},\ \bibinfo {pages} {085138}
  (\bibinfo {year} {2015})}\BibitemShut {NoStop}%
\end{thebibliography}%

\end{document}